\documentclass[12pt]{article}
\pdfoutput=1
\usepackage[nosort]{cite}

\usepackage{xcolor}

\usepackage{epsfig}
\usepackage{amsfonts}
\usepackage{amscd}
\usepackage{latexsym}
\usepackage{amsmath,amssymb}
\usepackage{verbatim}
\usepackage{setspace}
\usepackage{color}
\usepackage{fancyhdr}
\usepackage{cite}
\usepackage{hyperref}
\usepackage{tikz}
\usepackage{tikz-cd}
\usepackage{slashed}
\usepackage{soul}
\usepackage{multirow}
\usetikzlibrary{calc}
\usetikzlibrary{topaths}
\usetikzlibrary{decorations}
\usetikzlibrary{decorations.pathmorphing}

\usepackage{draft}
\usepackage{hyperref}
\usepackage{graphicx,color,subfig}
\usepackage{cite}
\usepackage{mciteplus}
\usepackage{skak}

\numberwithin{equation}{section}

\renewcommand{\b}[1]{\boldsymbol{#1}}
\newcommand{\diag}{\operatorname{diag}}
\newcommand{\im}{\operatorname{im}}
\newcommand{\coker}{\operatorname{coker}}

\newcommand{\Pic}{\operatorname{Pic}}
\newcommand{\Jac}{\operatorname{Jac}}

\def\({\left(}
\def\){\right)}

\begin{document}

\begin{titlepage}
	
\begin{center}

\hfill MIT-CTP/5450, YITP-SB-2022-24\\
\hfill \\
\vskip 1cm

 \title{Fractons on Graphs and Complexity}

\author{Pranay Gorantla$^{1}$, Ho Tat Lam$^{2}$, and Shu-Heng Shao$^{3}$}

\address{${}^{1}$Physics Department, Princeton University}
\address{${}^{2}$Center for Theoretical Physics, Massachusetts Institute of Technology}
\address{${}^{3}$C.\ N.\ Yang Institute for Theoretical Physics, Stony Brook University}

\end{center}

\vspace{2.0cm}

\begin{abstract}\noindent
We introduce two exotic lattice models on a general spatial graph. 
The first one is a matter theory of a compact Lifshitz scalar field, while the second one is a certain rank-2 $U(1)$ gauge theory of fractons. 
 Both lattice models are defined via the discrete Laplacian operator on a general graph. 
We unveil an intriguing correspondence between the physical observables of these lattice models and graph theory quantities. 
For instance, the ground state degeneracy of the matter theory equals the number of spanning trees of the spatial graph, which is a common measure of complexity  in graph theory (``GSD = complexity").
The discrete global symmetry is identified as the Jacobian group of the graph. 
In the gauge theory, superselection sectors of fractons are in one-to-one correspondence with the  divisor classes in graph theory. 
In particular, under mild assumptions on the spatial graph, the fracton immobility is proven using a graph-theoretic Abel-Jacobi map.

\end{abstract}

\vfill
	
\end{titlepage}

\eject

\tableofcontents

\section{Introduction}\label{sec:intro}

The past decades have seen an explosion of various exotic lattice models including gapped fracton models \cite{Chamon:2004lew,Haah:2011drr,Vijay:2016phm} (see \cite{Nandkishore:2018sel,Pretko:2020cko,Grosvenor:2021hkn} for reviews)\footnote{The word ``fracton'' was also used in other different contexts, for example \cite{Khlopov:1981aa,Alexander1982}.} and gapless models \cite{PhysRevB.66.054526} with subsystem symmetries.  
These models have various peculiar properties:
\begin{enumerate}
\item Exotic global symmetries such as (planar or fractal) subsystem global symmetries \cite{PhysRevB.66.054526,Vijay:2016phm}, multipole global symmetries \cite{Pretko:2016kxt,Pretko:2016lgv,Pretko:2018jbi,Gromov:2018nbv}, etc. (See \cite{McGreevy:2022oyu,Cordova:2022ruw} for recent reviews on generalized global symmetries \cite{Gaiotto:2014kfa}.)
\item In gapped fracton models, the logarithm of the ground state degeneracy (GSD)   grows, typically subextensively, with the linear size of the system \cite{Haah:2020ghp}.
\item Massive particle-like excitations that have restricted mobility---a particle can be completely immobile, a.k.a. \emph{fracton}, or can move only along a line, a.k.a. \emph{lineon}, etc.
\end{enumerate}
These peculiarities do not fit into the framework of conventional continuum quantum field theory. 
In particular, the gapped fracton models do not admit a  topological quantum field theory (TQFT) description at low energies. 
Instead, one has to go beyond the standard relativistic continuum field theory to describe them (see, for example,  \cite{Brauner:2022rvf} for a recent review, and references therein).

While there is a plethora of exotic lattice models, in most cases, the spatial lattice is assumed to be cubic with manifest translation invariance in the three spatial directions. There is usually no obvious way to define them on a triangulation of an arbitrary spatial manifold. See \cite{Gromov:2017vir,Slagle:2018kqf,Jain:2021ibh} for fracton models on more general manifolds. In some examples, a \emph{foliation} of spatial manifold is essential \cite{Slagle:2018wyl,Shirley:2017suz,Shirley:2018nhn,Shirley:2018hkm,Shirley:2018vtc,Slagle:2018swq,
Shirley:2019uou,Slagle:2020ugk,Hsin:2021mjn,Geng:2021cmq}. In contrast, standard lattice models, such as the Ising model, or the toric code, can be defined on an arbitrary triangulation of the spatial manifold.

What is the minimal structure we need to assume about the lattice? We need a set of vertices, which host the degrees of freedom, and a set of edges connecting them. This defines a mathematical object known as a \emph{graph}. For example, the quantum Ising model can be defined on any spatial graph where the interaction is along the edges. In fact, lattice models on general graphs can be engineered in cold atom experiments. See \cite{Bentsen:2019bra,Bentsen:2019rlr,Periwal:2021eur} for examples. One is then naturally led to the question: \emph{can we construct exotic lattice models such as fractons on a general graph?} (See \cite{Manoj:2021rpq} for an example in this direction.)

In this work, we propose two lattice models on an arbitrary finite, undirected graph $\Gamma$. (For simplicity, we assume that the graph is simple and connected. See Section \ref{sec:graph} for the definitions of these adjectives.) We work with a Euclidean spacetime where each spatial slice is $\Gamma$ (see Figure \ref{fig:spacetime}).  Our results can easily be recast in a Hamiltonian formulation. One model is a matter theory based on a compact scalar field $\phi$, while the other is a pure $U(1)$ gauge theory associated with the global symmetry of the matter theory. We refer to them as the \emph{Laplacian $\phi$-theory} and the \emph{$U(1)$ Laplacian gauge theory}, respectively, because they are constructed using the \emph{discrete Laplacian operator} $\Delta_L$ on a general graph $\Gamma$.

Na\"ively, the Laplacian $\phi$-theory can be viewed as a particular regularization of the compact scalar Lifshitz theory described by the Lagrangian
\ie
\mathcal L = \frac{\mu_0}{2} (\partial_\tau \phi)^2 + \frac{\mu}{2} (\nabla^2 \phi)^2~,
\fe
where $\nabla^2$ is the Laplacian differential operator on the spatial manifold. On the other hand, the $U(1)$ Laplacian gauge theory can be viewed as a particular regularization of a rank-2 $U(1)$ gauge theory with gauge fields $(A_\tau,A)$ satisfying the gauge symmetry\footnote{We call this a ``rank-2'' gauge theory because the gauge transformation of the spatial gauge field involves a second-order spatial derivative.}
\ie
A_\tau \sim A_\tau + \partial_\tau \alpha~, \qquad A \sim A + \nabla^2 \alpha~.
\fe
The Lagrangian is
\ie
\mathcal L = \frac{1}{2g^2} E^2~,
\fe
where $E = \partial_\tau A - \nabla^2 A_\tau$ is the gauge invariant electric field. 
However, these continuum Lagrangians do not specify the systems unambiguously. 
In this paper, we use the modified Villain formulation  \cite{Sulejmanpasic:2019ytl,Gorantla:2021svj} to provide a precise formulation of these systems.  
We will see that various physical observables (including the GSD) depend sensitively on how the space is discretized by a discrete lattice graph $\Gamma$.  
In particular, the discrete Laplacian difference operator $\Delta_L$ does not have  a smooth continuum limit to the differential operator $\nabla^2$.

It is important to emphasize that the Laplacian $\phi$-theory is not robust. Small deformations of the short distance theory change the elaborate long distance structure that we find here. (See the discussion in \cite{paper1}.)  On the other hand, the $U(1)$ Laplacian gauge theory is a robust model. In an upcoming paper \cite{Gorantla:2022pii}, we will discuss gapped robust models that are related to these models.

Since our models can be placed on any spatial graph $\Gamma$, they can be defined in general spatial dimensions. In particular, we can take the graph $\Gamma$ to be a $d$-dimensional torus lattice for any $d\ge1$. We will examine our models on 1d and 2d torus graphs in details below. More generally, we can place our models on a general graph where there is no clear notion of dimensionality or locality.  In particular, foliation structure is not needed to define these models.

In analyzing these models, we follow the approach advocated in \cite{Seiberg:2019vrp,paper1,paper2,paper3,Gorantla:2020xap,Gorantla:2020jpy,Rudelius:2020kta,Gorantla:2021svj,Gorantla:2021bda,Burnell:2021reh,Gorantla:2022eem}, i.e., we focus on the global symmetries and pursue their consequences. Here are the highlights of these models:
\begin{enumerate}
\item The discrete global symmetry of the matter theory is based on the \emph{Jacobian group} of the graph $\Gamma$, denoted as $\Jac(\Gamma)$, which is a well-studied finite abelian group associated with a general graph in graph theory.  Relatedly,  the discrete time-like global symmetry \cite{Gorantla:2022eem}, which acts on  defects, is also given by $\Jac(\Gamma)$ in the gauge theory.
\item In the matter theory, the ground state degeneracy  is equal to $|\Jac(\Gamma)|$, the order of the Jacobian group. This in turn is equal to the number of \emph{spanning trees} of $\Gamma$, which we will define in the main text.
The number of spanning trees is a common measure of the \emph{complexity} of a general graph in graph theory (see, for example, \cite{GRONE198897,alon90}). 
Therefore, at the level of a slogan, we have 
\ie
\text{GSD = complexity}~.
\fe
This is one manifestation of the UV/IR mixing phenomenon observed in many of these exotic models: certain low-energy/long-distance observables depend sensitively on the short-distance details \cite{paper1,Gorantla:2021bda}. 
In our matter theory, the GSD equals  the complexity of  the discretized spatial graph, which  can be thought of as a short-distance regularization. 
Under a mild assumption on $\Gamma$, the ground state degeneracy grows exponentially in the number of vertices of $\Gamma$ \cite{alon90}.\footnote{One might wonder if this system is as trivial as decoupled spins on the sites of the graph because the GSD grows exponentially in $\mathsf N$ in both cases. However, as we will show below, the origins of this exponential behavior are very different. When both systems are placed on a 2d torus spatial lattice with $L$ sites in each direction, the minimal number of generators of discrete momentum symmetry group of the Laplacian $\phi$-theory grows only linearly in $L$, whereas it grows as $L^2$ in the decoupled spin system. So the large GSD of the Laplacian $\phi$-theory comes from the large orders of some of the generators of $\Jac(\Gamma)$ rather than the number of generators.} 
\item  In the gauge theory, the  defects, which describe the worldlines of infinitely heavy probe particles,   are immobile under a mild assumption on the graph $\Gamma$. Therefore, the $U(1)$ Laplacian gauge theory is  a  theory of \emph{fractons on a general graph}.\footnote{In contrast to the matter theory, the ground state of the gauge theory (which has fractons) is non-degenerate (assuming that the $\theta$-angle is not $\pi$). See Section \ref{sec:U1spectrum}.} There is a beautiful correspondence between the physical observables for fractons and various graph-theoretic concepts. In particular, the space of superselection sectors for fractons is translated into the theory of divisors of a graph.
\end{enumerate}
These features are analogous to those at the beginning of this introduction. Interestingly, they are intimately related to some well-studied properties of a graph.

This paper is organized as follows. In Section \ref{sec:graph}, we collect some useful mathematical facts about graphs, and functions on their vertices. We define the Laplace difference operator $\Delta_L$ on the graph $\Gamma$, and discuss the properties of solutions to discrete Laplace and Poisson equations. We introduce the theory of divisors on a finite graph, and define the Picard group and the Jacobian group of $\Gamma$. We also define the Abel-Jacobi map on the graph and discuss its properties. We relate the Jacobian group to the Smith normal form of the Laplacian operator and the spanning trees of $\Gamma$. These results can be found in any standard textbook on graph theory such as \cite{Chung:97,corry2018divisors}. In Appendix \ref{app:lap-poisson}, we solve the discrete Poisson equation on any graph using the Smith decomposition of the Laplacian matrix.

In Section \ref{sec:lapphi}, we introduce the Laplacian $\phi$-theory on the graph $\Gamma$. It is a self-dual model with momentum and winding symmetries. The momentum (or winding) symmetry is $U(1) \times \Jac(\Gamma)$, where $\Jac(\Gamma)$ is the Jacobian group of $\Gamma$. The noncommutativity of the momentum and winding symmetries leads to a ground state degeneracy equal to $|\Jac(\Gamma)|$, the order of the Jacobian group. This is also equal to the number of spanning trees or the complexity of $\Gamma$. We note that this model is not robust---deforming the theory by a winding operator breaks the winding symmetry and lifts the degeneracy.

In Section \ref{sec:lapA}, we introduce the $U(1)$ Laplacian gauge theory on the graph $\Gamma$. It is the pure gauge theory associated with the momentum symmetry of the Laplacian $\phi$-theory. While the space-like global symmetry of this model is simply $U(1)$, the time-like global symmetry is $U(1) \times \Jac(\Gamma)$. This leads to selection rules on the mobility of defects. In particular, we prove that a defect with unit charge is completely immobile when $\Gamma$ is $2$-edge connected (we will define this below). In other words, a single charged particle is a fracton. We also give a complete characterization of mobility of defects for arbitrary $\Gamma$. Finally, we note an interesting correspondence between divisors in graph theory and configurations of fractons on $\Gamma$, which is summarized in Table \ref{tbl:div_fracton}.

\renewcommand{\arraystretch}{1.5}
\begin{table}
\begin{center}
\begin{tabular}{|c|c|}
\hline
Theory of divisors & $U(1)$ Laplacian gauge theory
\tabularnewline
\hline
\hline
Graph $\Gamma$ & Spatial lattice
\tabularnewline
\hline
Divisor & Configuration of fractons with
\tabularnewline
$q\in\mathcal F(\Gamma,\mathbb Z)$ & $U(1)$ time-like charges $q(i)$ at site $i$
\tabularnewline
\hline
Principal divisor & Configuration of fractons in
\tabularnewline
$q\in\im_{\mathbb Z} \Delta_L$ & trivial superselection sector
\tabularnewline
\hline
Degree of $q$ & Total $U(1)$ time-like charge of 
\tabularnewline
$\deg q := \sum_i q(i)$ & a configuration of fractons
\tabularnewline
\hline
Picard group & Space of all superselection sectors, or
\tabularnewline
$\Pic(\Gamma)$ & space of all time-like charges
\tabularnewline
\hline
Jacobian group & Space of superselection sectors
\tabularnewline
$\Jac(\Gamma)$ & with trivial total $U(1)$ time-like charge
\tabularnewline
\hline
Pontryagin dual of $\Pic(\Gamma)$ & \multirow{2}{*}{Time-like symmetry group}
\tabularnewline
$U(1)\times\Jac(\Gamma)$ &
\tabularnewline
\hline
\end{tabular}
\end{center}
\caption{The correspondence between the theory of divisors on the graph $\Gamma$, and the $U(1)$ Laplacian gauge theory on the spatial lattice $\Gamma$. The graph-theoretic objects in the left column are defined in Section \ref{sec:graph}, and the physical observables in the right column are discussed in Section \ref{sec:lapA}.}\label{tbl:div_fracton}
\end{table}

In both Sections \ref{sec:lapphi} and \ref{sec:lapA}, we discuss two concrete examples where we place each model on spatial circle and torus respectively. While the former gives the 1+1d dipole theories analyzed in \cite{Gorantla:2022eem}, the latter gives new 2+1d models with interesting properties. These new 2+1d models can be interpreted as extensions of the 1+1d dipole theories of \cite{Gorantla:2022eem} to 2+1d. In fact, there are other ways to extend the 1+1d dipole theories to 2+1d. In an upcoming paper \cite{Gorantla:2022ssr}, we compare all these 2+1d models, and discuss their relation to the 2+1d compact Lifshitz theory \cite{Henley1997,Moessner2001,Vishwanath:2004,Fradkin:2004,Ardonne:2003wa,Ghaemi2005,Chen:2009ka,2018PhRvB..98l5105M,Yuan:2019geh,Lake:2022ico} and 2+1d rank-2 $U(1)$ tensor gauge theory \cite{Pretko:2016kxt,Pretko:2016lgv}.

In another upcoming paper \cite{Gorantla:2022pii}, we propose and analyze two gapped $\mathbb Z_N$ models on a graph. One of them is a fracton model, which is a Higgsed version of the $U(1)$ Laplacian gauge theory on the graph. The other is a robust lineon model. We also study these gapped models on a spatial torus, and compare them with 2+1d rank-2 $\mathbb Z_N$ tensor gauge theory \cite{Bulmash:2018lid,Ma:2018nhd,Oh2021,Oh2022,Pace2022}.

\section{Graph theory primer}\label{sec:graph}
In this section, we collect some important mathematical facts about a finite graph, and functions on the graph valued in abelian groups such as $\mathbb R$, $U(1)$, $\mathbb Z$, or $\mathbb Z_N$. Most of the details can be found in a standard textbook on spectral graph theory such as \cite{Chung:97}. The theory of divisors on finite graphs is discussed in \cite{corry2018divisors}.

Let $\Gamma$ be a \emph{simple} (at most one edge between any two vertices and no self-loops), \emph{undirected} (no directed edges), \emph{connected} (any two vertices are connected by a path) graph on $\mathsf N$ vertices. We use $i$ to denote a vertex (or site), and $\langle i,j\rangle$ or $e$ to denote an edge (or link).

The \emph{adjacency matrix} $A$ of $\Gamma$ is an $\mathsf N\times \mathsf N$ symmetric matrix given by $A_{ij} = 1$ if there is an edge $\langle i,j\rangle$ between vertices $i$ and $j$, and $A_{ij}=0$ otherwise. The \emph{degree} $d_i$ of a vertex $i$ is the number of edges incident to the vertex $i$. Let $D = \diag(d_1,\ldots,d_{\mathsf N})$ be the degree matrix. The \emph{Laplacian matrix} $L$ of $\Gamma$ is defined as $L:=D-A$. Note that $L$ is symmetric.

Here are some common examples/classes of graphs:
\begin{itemize}
\item A \emph{$k$-regular graph} is a graph where every vertex has degree $k$.
\item A \emph{$k$-edge connected graph} is a graph where removing any $k-1$ edges still leaves it connected.
\item The \emph{complete graph} on $\mathsf N$ vertices, denoted as $K_{\mathsf N}$, is a graph that has an edge between any two vertices. Equivalently, it is the only $(\mathsf N-1)$-regular graph on $\mathsf N$ vertices.
\item The \emph{cycle graph} on $\mathsf N$ vertices, denoted as $C_{\mathsf N}$, is a graph that is a cycle or loop. Equivalently, it is the only $2$-regular graph on $\mathsf N$ vertices.
\item A \emph{tree} on $\mathsf N$ vertices is a graph that contains no cycle or loop. If it is connected, then it is called a \emph{spanning tree}.
\end{itemize}

\subsection{Discrete harmonic functions and Smith decomposition of $L$}

Consider a function on the vertices of the graph, $f:\Gamma\rightarrow X$, where $X$ is an abelian group. We denote the set of all such functions as $\mathcal F(\Gamma,X)$. Define the \emph{discrete Laplacian} operator $\Delta_L:\mathcal F(\Gamma,X)\rightarrow \mathcal F(\Gamma,X)$ as
\ie
\Delta_L f(i) := \sum_j L_{ij} f(j) = d_i f(i) - \sum_{j:\langle i,j\rangle\in \Gamma} f(j) = \sum_{j:\langle i,j\rangle\in \Gamma} [f(i) - f(j)]~.
\fe
The additions and subtractions here are with respect to the group multiplication of $X$. This is one of the most natural and universal difference operators that can be defined on any such graph $\Gamma$. The \emph{image} of $\mathcal F(\Gamma,X)$ under $\Delta_L$ is denoted as $\im_X\Delta_L$. A function $f\in\mathcal F(\Gamma,X)$ is said to be \emph{discrete harmonic} if it satisfies
\ie
\Delta_L f(i) = 0~,
\fe
where $0$ is the identity element in $X$. We denote the set of all $X$-valued discrete harmonic functions as $\mathcal H(\Gamma,X)$, or $\ker_X\Delta_L$, the \emph{kernel} of $\Delta_L$.\footnote{The space $\ker_X \Delta_L$ is also known as the \emph{group of balanced vertex weightings} \cite{Berman:1986}.}

 Given a $g\in\mathcal F(\Gamma,X)$, consider the \emph{discrete Poisson equation}  
\ie\label{graph-discpoiss}
\Delta_L f(i) = g(i)~.
\fe
If $g=0$, then \eqref{graph-discpoiss} is called a \emph{discrete Laplace equation}. We define the \emph{cokernel} of $\Delta_L$ as the quotient
\ie
\coker_X \Delta_L := \frac{\mathcal F(\Gamma,X)}{\im_X \Delta_L}~,
\fe
Trivially, a solution to the discrete Poisson equation \eqref{graph-discpoiss} exists if and only if $g$ is in the same equivalence class as $0$ in this quotient.

Solutions to the discrete Poisson equation \eqref{graph-discpoiss}, if any, can be found using the \emph{Smith decomposition} \cite{Smith:1861} of the Laplacian matrix $L$ \cite{Lorenzini:08}. The \emph{Smith normal form} of $L$ is given by
\ie\label{graph-snf}
R = P L Q~,\qquad \text{or}\qquad R_{ab} = \sum_{i,j} P_{ai} L_{ij} Q_{jb}~,
\fe
where $P,Q\in GL_{\mathsf N}(\mathbb Z)$, and $R = \diag(r_1,\ldots,r_{\mathsf N})$ is an $\mathsf N\times \mathsf N$ diagonal integer matrix with nonnegative diagonal entries, known as the \emph{invariant factors} of $L$, such that $r_a | r_{a+1}$ (i.e., $r_a$ divides $r_{a+1}$) for $a=1,\ldots, \mathsf N-1$. While $R$ is uniquely determined by $L$, the matrices $P$ and $Q$ are not. More details on the structure of $P$ and $Q$, and how to solve \eqref{graph-discpoiss} using the Smith decomposition of $L$ can be found in Appendix \ref{app:lap-poisson}.

One important result that we will repeatedly use is the general solution to the $U(1)$-valued discrete Laplace equation. In Appendix \ref{app:lap-poisson} (see in particular \eqref{Poisson-sol-U1}), we show that the most general $U(1)$-valued discrete harmonic function $f\in\mathcal{H}(\Gamma,U(1))$ on a graph $\Gamma$ is given by
\ie\label{Laplace-sol-U1}
f(i) =  2\pi \sum_{a< \mathsf N} \frac{Q_{ia} p_a}{r_a} + c \in \mathbb{R}/2\pi\mathbb{Z}~,
\fe
parametrized by a circle-valued constant $c$, i.e., $c\sim c+2\pi$, and $\mathsf{N}-1$ integers $p_a = 0,1,\ldots, r_a-1$. If we lift this solution to $\mathbb{R}$, then it obeys
\ie\label{Laplace-sol-U1-int}
\Delta_L f(i) =  2\pi \sum_{a<\mathsf N} (P^{-1})_{ia} p_a\in 2\pi \mathbb{Z}~.
\fe

\subsection{Theory of divisors and the Abel-Jacobi map on a graph}

There is an interesting analogy between the theory of integer-valued functions on a finite graph and the theory of divisors on a Riemann surface \cite{BSMF:97,BAKER2007,corry2018divisors}. In this context, an element of $\mathcal F(\Gamma,\mathbb Z)$ is known as a \emph{divisor}, and an element of $\im_{\mathbb Z}\Delta_L$ is known as a \emph{principal divisor}. Given a divisor $q$, its \emph{degree} is defined as $\deg q := \sum_i q(i)$. Let $\mathcal F^k(\Gamma,\mathbb Z)$ denote the set of all degree-$k$ divisors. Note that any principal divisor has degree zero, so we can define the quotients:
\ie
\Pic(\Gamma) := \frac{\mathcal F(\Gamma,\mathbb Z)}{\im_{\mathbb Z}\Delta_L} = \coker_{\mathbb Z}(\Delta_L)~,\qquad \Jac(\Gamma) := \frac{\mathcal F^0(\Gamma,\mathbb Z)}{\im_{\mathbb Z}\Delta_L}~,
\fe
known as the \emph{Picard group}, and the \emph{Jacobian group} respectively. As the names suggest, they are groups; in fact, they are abelian groups. They are related by the split-exact sequence
\ie
0 \longrightarrow \Jac(\Gamma) \longrightarrow \Pic(\Gamma) \xrightarrow{\,\deg\,} \mathbb Z \longrightarrow 0~.
\fe

 The \emph{characteristic function} of a vertex $i$ is defined as $\chi_i(j) := \delta_{ij}$. Given two vertices $i,i'\in\Gamma$, we define the function $s_{i,i'}(j) := \chi_i(j) - \chi_{i'}(j)$. For a fixed vertex $i_0\in\Gamma$, the \emph{Abel-Jacobi map}, $S_{i_0}:\Gamma \rightarrow \Jac(\Gamma)$, is defined as
\ie\label{graph-AJmap}
S_{i_0}(i) := [s_{i,i_0}]~.
\fe
It is well-defined because $s_{i,i_0} \in \mathcal F^0(\Gamma,\mathbb Z)$. Even though $i$ and $i_0$ play similar roles in the right-hand side of this equation, they will play different roles below. This is the reason for the asymmetry between them in the left-hand side of \eqref{graph-AJmap}.

The Abel-Jacobi map enjoys several nice properties:
\begin{enumerate}
\item It is a $\Jac(\Gamma)$-valued discrete harmonic function that vanishes at $i_0$.

\item For any abelian group $X$, given a group homomorphism $\psi:\Jac(\Gamma)\rightarrow X$, the composition $\psi \circ S_{i_0}:\Gamma \rightarrow X$ is an $X$-valued discrete harmonic function that vanishes at $i_0$. Conversely, for any $X$-valued discrete harmonic function $f$ that vanishes at $i_0$, there is a unique group homomorphism $\psi_f:\Jac(\Gamma)\rightarrow X$ such that $f = \psi_f \circ S_{i_0}$. This is known as the \emph{universal property} of the Abel-Jacobi map.

\item It is injective if and only if $\Gamma$ is $2$-edge connected. More generally, for any two vertices $i,i'\in\Gamma$, $S_{i_0}(i) = S_{i_0}(i')$ if and only if there is a unique path from $i$ to $i'$ in $\Gamma$.
\end{enumerate}
These properties of the Abel-Jacobi map turn the study of discrete harmonic functions on a graph to a problem in group theory.

\subsection{Jacobian group and complexity of a graph}

The Jacobian group $\Jac(\Gamma)$ defined in the previous subsection is a natural finite abelian group that is associated with a general graph $\Gamma$.\footnote{It has several different names in the graph theory literature, including the \emph{sandpile group} \cite{Dhar:90,Dhar_1995}, or the \emph{group of components} \cite{Lorenzini:91}, or the \emph{critical group} \cite{Biggs:1999vy} of $\Gamma$, and it is related to the \emph{group of bicycles} \cite{Berman:1986} of $\Gamma$. In particular, it would be interesting to understand the connection between our models and the abelian sandpile model.} It is closely related to the Smith normal form of the Laplacian matrix $L$ (see Appendix \ref{app:lap-poisson}, especially, around \eqref{gaugefix-Z}). In particular, we have
\ie
\Jac(\Gamma) \cong \prod_{a<\mathsf N} \mathbb Z_{r_a}~.
\fe

The order of $\Jac(\Gamma)$ can be expressed in terms of   the nonzero eigenvalues of $L$:
\ie\label{graph-matrixtree}
|\Jac(\Gamma)| = \prod_{a<\mathsf N} r_a = \frac{\lambda_2 \cdots \lambda_{\mathsf N}}{\mathsf N}~,
\fe
where $0 = \lambda_1 < \lambda_2 \le \cdots \le \lambda_{\mathsf N}$ are the eigenvalues of $L$.\footnote{The eigenvalues of $L$ are real because $L$ is symmetric. The zero eigenvalue of $L$ corresponds to the zero mode of the Laplacian operator $\Delta_L$. The other eigenvalues are all positive because $\Gamma$ is connected.} By  \emph{Kirchhoff's matrix-tree theorem} \cite{Kirchhoff:1847}, this is equal to the number of \emph{spanning trees} of $\Gamma$. Here, a spanning tree of $\Gamma$ is a subgraph that is a spanning tree on the vertices of $\Gamma$. See Figure \ref{fig:spacetime}(b) for an example.

The number of spanning trees is the most fundamental and well-studied notion of \textit{complexity} in graph theory. Intuitively, it is a measure of how ``connected'' $\Gamma$ is. For example, it is easy to see that the number of spanning trees of $C_{\mathsf N}$, the cycle graph on $\mathsf N$ vertices, is $\mathsf N$. In contrast, an old result of Cayley states that the number of spanning trees of $K_{\mathsf N}$, the complete graph on $\mathsf N$ vertices, is $\mathsf N^{\mathsf N-2}$ \cite{cayley1889}.\footnote{It is not too difficult to prove this using \eqref{graph-matrixtree}.} More generally, when $\Gamma$ is $k$-regular, the number of spanning trees of $\Gamma$ grows exponentially in $\mathsf N$ whenever $k\ge 3$ \cite{alon90}.

\section{Laplacian $\phi$-theory on a graph}\label{sec:lapphi}

\begin{figure}[t]
\begin{center}
~\hfill\includegraphics[scale=0.2]{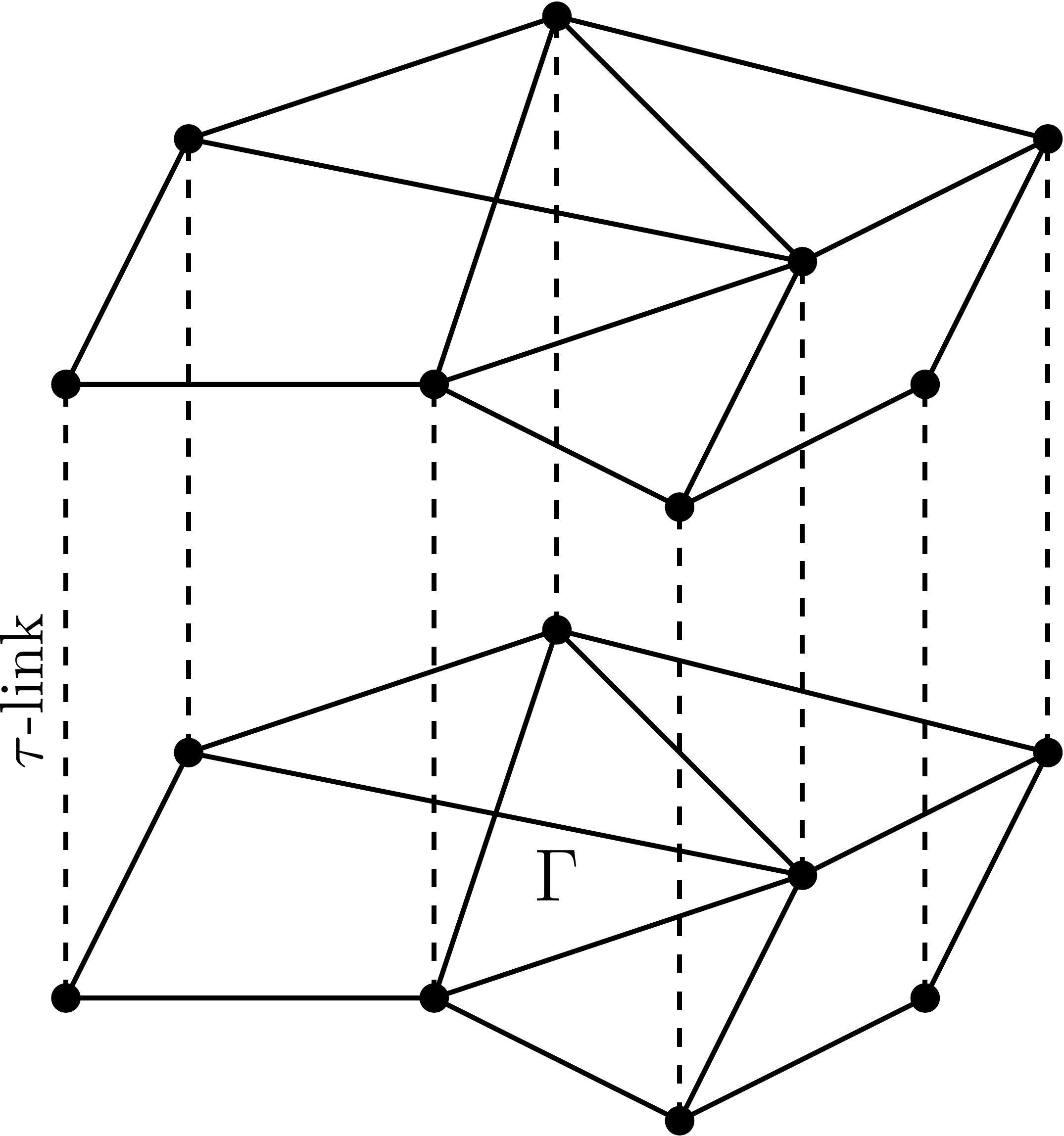}\hfill
\raisebox{0.2\height}{\includegraphics[scale=0.3]{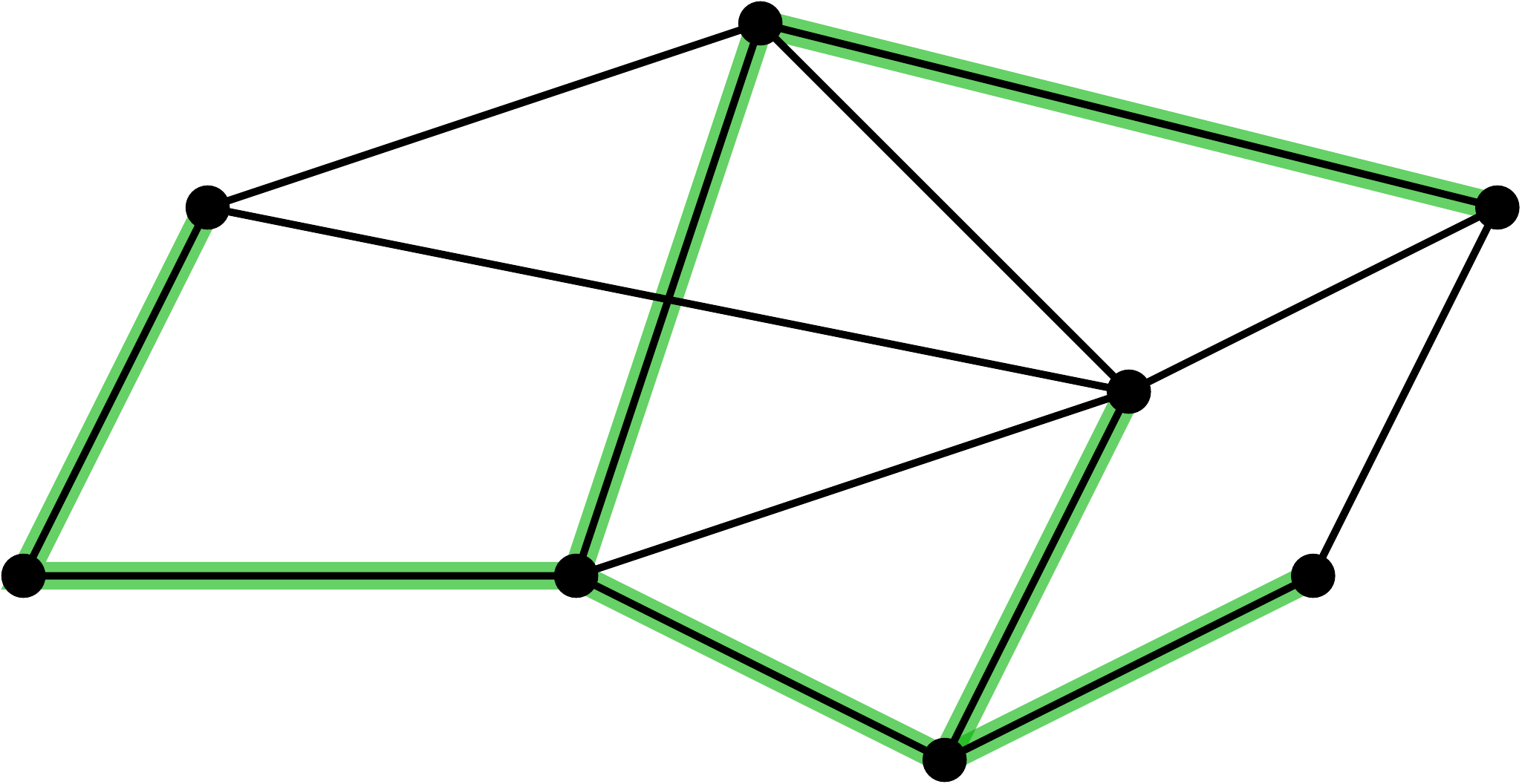}}\hfill~
\\
\hfill (a)~~~~~~~~~~ \hfill ~~~~~~(b) \hfill~~~~~
\end{center}
\caption{(a) The Euclidean spacetime $\mathbb Z_{L_\tau}\times \Gamma$: the solid lines and dots correspond to a spatial slice which is the graph $\Gamma$, and the dashed lines represent the $\tau$-links. (b) A spanning tree of $\Gamma$ associated with the highlighted (green) edges.}\label{fig:spacetime}
\end{figure}

In this section, we study the modified Villain version \cite{Gorantla:2021svj} of the Laplacian $\phi$-theory on a graph $\Gamma$.  
 It can be viewed as an extension of the 1+1d dipole $\phi$-theory of \cite{Gorantla:2022eem} with $\Delta_x^2$ on the 1d spatial lattice replaced by the discrete Laplacian operator $\Delta_L$ on the graph $\Gamma$. The Euclidean spacetime is $\mathbb Z_{L_\tau} \times \Gamma$, i.e., each spatial slice is $\Gamma$, and any two adjacent spatial slices are connected by $\tau$-links at each vertex (see Figure \ref{fig:spacetime}(a)).

The modified Villain version of the Laplacian $\phi$-theory is described by the action\footnote{One can also study the \emph{Laplacian XY model}, which is described by the following  action:
\ie
S = -\beta_0 \sum_{\hat \tau,i} \cos\left[ \Delta_\tau \varphi(\hat \tau,i) \right] - \beta \sum_{\hat \tau,i} \cos\left[ \Delta_L \varphi(\hat \tau,i) \right]~,
\fe
where $\varphi(\hat \tau,i)$ is a circle-valued field on each site of the spacetime lattice. The momentum symmetry of this cosine model is the same as that of the modified Villain model \eqref{lapphi-modVill-action}.}
\ie\label{lapphi-modVill-action}
S &= \frac{\beta_0}{2} \sum_{\hat \tau,i} \left[\Delta_\tau \phi(\hat \tau,i)- 2\pi n_\tau(\hat \tau,i) \right]^2 + \frac{\beta}{2} \sum_{\hat \tau,i} \left[ \Delta_L \phi(\hat \tau,i) - 2\pi n(\hat \tau,i) \right]^2
\\
& \qquad + i\sum_{\hat \tau,i} \tilde \phi(\hat \tau,i) \left[ \Delta_\tau n(\hat \tau,i) - \Delta_L n_\tau(\hat \tau,i) \right]~,
\fe
where $\phi$ is a real-valued field at each site, $n$ is an integer gauge field at each site, $n_\tau$ is an integer gauge field on each $\tau$-link, and $\tilde \phi$ is a real-valued Lagrange multiplier on each $\tau$-link.\footnote{We associate the $\tau$-link between $(\hat \tau,i)$ and $(\hat \tau+1,i)$ to the site $(\hat \tau,i)$.} 
Recall that $\sum_i$ stands for the sum over all the sites $i$ of the graph $\Gamma$.  
There is a gauge symmetry
\ie\label{lapphi-gaugesym}
&\phi(\hat \tau,i) \sim \phi(\hat \tau,i) + 2\pi k(\hat \tau,i)~,
\\
&n_\tau(\hat \tau,i) \sim n_\tau(\hat \tau,i) + \Delta_\tau k(\hat \tau,i)~,
\\
&n(\hat \tau,i) \sim n(\hat \tau,i) + \Delta_L k(\hat \tau,i)~,
\\
&\tilde \phi(\hat \tau,i) \sim \tilde \phi(\hat \tau,i) + 2\pi \tilde k(\hat \tau,i)~,
\fe
where $k$ and $\tilde k$ are integer gauge parameters on the sites and $\tau$-links, respectively. This integer gauge symmetry makes the scalar fields $\phi$ and $\tilde \phi$ compact. The lattice action \eqref{lapphi-modVill-action} is a particular lattice regularization of a compact Lifshitz scalar field theory that can be defined on a general graph $\Gamma$.  See \cite{Gorantla:2021svj} for similar modified Villain formulations of various standard and exotic theories of compact scalar fields.

\subsection{Self-duality}
Since $L$ is symmetric, the modified Villain model \eqref{lapphi-modVill-action} is self-dual with $\phi \leftrightarrow \tilde \phi$ and $\beta_0 \leftrightarrow \frac{1}{(2\pi)^2\beta}$. Indeed, using the Poisson resummation formula for the integers $n_\tau,n$, the dual action is
\ie
S &= \frac{1}{2(2\pi)^2\beta} \sum_{\hat \tau,i} \left[\Delta_\tau \tilde \phi(\hat \tau,i)- 2\pi \tilde n_\tau(\hat \tau,i) \right]^2 + \frac{1}{2(2\pi)^2\beta_0} \sum_{\hat \tau,i} \left[ \Delta_L \tilde \phi(\hat \tau,i) - 2\pi \tilde n(\hat \tau,i) \right]^2
\\
& \qquad - i\sum_{\hat \tau,i} \phi(\hat \tau,i) \left[ \Delta_\tau \tilde n(\hat \tau,i) - \Delta_L \tilde n_\tau(\hat \tau,i) \right]~,
\fe
where $(\tilde n_\tau,\tilde n)$ are integer gauge fields that make $\tilde \phi$ compact. Under the gauge symmetry \eqref{lapphi-gaugesym}, they transform as
\ie
\tilde n_\tau(\hat \tau,i) \sim \tilde n_\tau(\hat \tau,i) + \Delta_\tau \tilde k(\hat \tau,i)~,\qquad \tilde n(\hat \tau,i) \sim \tilde n(\hat \tau,i) + \Delta_L \tilde k(\hat \tau,i)~.
\fe
When the graph is a two-dimensional torus graph, this is related to the self-duality of the 2+1d compact Lifshitz scalar field theory discussed in \cite{Vishwanath:2004}.

\subsection{Global symmetry and the Jacobian group}
The momentum global symmetry of the action \eqref{lapphi-modVill-action} corresponds to shifting the fields $\phi,n$ by
\ie
&\phi (\hat\tau ,i ) \to \phi(\hat\tau, i ) + f(i)~,
\\
&n(\hat\tau , i ) \to n(\hat \tau, i) + \frac1{2\pi} \Delta_L f(i)~,
\fe
where the function $f(i)$ obeys $\Delta_L f(i ) \in 2\pi \mathbb{Z}$. In other words, $f(i)$  is a solution to the $U(1)$-valued discrete Laplace equation, i.e., $f\in \mathcal{H}(\Gamma,U(1))$. The simplest example of such an $f(i)$ is a constant, i.e., $f(i) = c$ with $c\sim c+2\pi$. This corresponds to a $U(1)$ momentum symmetry. 

The most general solution to the $U(1)$-valued discrete Laplace equation is given in \eqref{Laplace-sol-U1}, which leads to the following momentum symmetry
\ie\label{lapphi-momsym}
&\phi(\hat \tau, i) \rightarrow \phi(\hat \tau,i) + 2\pi \sum_{a< \mathsf N} \frac{Q_{ia} p_a}{r_a} + c~,
\\
&n(\hat \tau,i) \rightarrow n(\hat \tau,i) + \sum_{a<\mathsf N} (P^{-1})_{ia} p_a~.
\fe
The momentum symmetry is parametrized by a circle-valued constant $c$ and  $\mathsf N-1$ integers $p_a = 0,\ldots,r_a -1$. 
Here $P$, $Q$ and $r_a$ are associated with the Smith decomposition of $L$. 
Note that the shift in $n$ is given by \eqref{Laplace-sol-U1-int}. 
The parameter   $p_a$ generates a $\mathbb Z_{r_a}$ discrete momentum symmetry for each $a<\mathsf N$. The total  momentum symmetry is therefore $U(1)\times \Jac(\Gamma)$ where
\ie\label{lapphi-dismomsym}
\Jac(\Gamma) = \prod_{a< \mathsf N} \mathbb Z_{r_a}~,
\fe
 is the Jacobian group of the graph $\Gamma$.  (See \cite{Berman:1986} for an alternative interpretation of the momentum symmetry group $U(1)\times \Jac(\Gamma)$.)  As we will see in Section \ref{sec:2dlapphi}, when $\Gamma$ is a 2d torus lattice $C_L \times C_L$, the minimal number of generators of $\Jac(\Gamma)$ grows only linearly in $L$.

In addition to the $U(1)\times \Jac(\Gamma)$ momentum symmetry, there is also a $U(1)\times \Jac(\Gamma)$ winding symmetry. This is to be expected given the self-duality of the theory. The $U(1)$ winding charge is
\ie\label{lapphi-modVill-U1wind-charge}
\tilde{\mathcal Q} = \sum_{i} P_{\mathsf N i} n(\hat \tau,i) = \sum_i n(\hat \tau,i)~,
\fe
while the $\mathbb Z_{r_a}$ discrete winding charge is
\ie\label{lapphi-modVill-discwind-charge}
\tilde{\mathcal Q}_a = \sum_i P_{ai} n(\hat \tau,i) \mod r_a~,
\fe
for each $a<\mathsf N$. They are conserved due to the flatness of $(n_\tau,n)$ imposed by the Lagrange multiplier $\tilde \phi$.

\subsection{GSD = Complexity}\label{sec:lapphi-gsd}

In this subsection, we compute the ground state degeneracy of the Laplacian $\phi$-theory. To facilitate this computation, we find it convenient to first gauge fix the integer gauge fields. First, we gauge fix $n_\tau = 0$ everywhere except at $\hat \tau = 0$. The remaining integer gauge freedom is the set of time-independent gauge transformations $k(i)$. By the flatness condition, we also have $\Delta_\tau n=0$, so $n(\hat \tau,i)=n(i)$. By the analysis in Appendix \ref{app:lap-poisson} around \eqref{gaugefix-Z}, the honolomies of $n(i)$ with gauge parameter $k(i)$ are precisely the winding charges \eqref{lapphi-modVill-U1wind-charge} and \eqref{lapphi-modVill-discwind-charge}.

In this gauge, a discrete winding configuration with discrete winding charges $\tilde{\mathcal Q}_a = p_a \mod r_a$ for $a<\mathsf N$ and zero $U(1)$ winding charge, $\tilde{\mathcal Q}=0$, is given by\footnote{This configuration has $\tilde{\mathcal Q} = 0$ because $\sum_i (P^{-1})_{ia}=0$ for $a<\mathsf N$ as shown in \eqref{Pinv-Qinv-sum}.}
\ie\label{lapphi-disc-wind-config}
\phi(\hat \tau,i)  = 2\pi \sum_{a< \mathsf N} \frac{Q_{ia} p_a}{r_a}~,\qquad n(\hat \tau,i) = \sum_{a<\mathsf N} (P^{-1})_{ia} p_a~.
\fe
This is the configuration in \eqref{Laplace-sol-U1} with $c=0$. There are $|\Jac(\Gamma)|=\prod_{a<\mathsf N} r_a$ such discrete winding configurations labeled by the $p_a$'s.  
All these configurations  have zero energy. Therefore, the ground state degeneracy is
\ie\label{lapphi-gsd}
\text{GSD} = |\Jac(\Gamma)| = \prod_{a<\mathsf N} r_a~.
\fe
As mentioned in Section \ref{sec:graph}, this is equal to the number of spanning trees of the graph $\Gamma$, which measures how complex a graph is. It follows that, when $\Gamma$ is $k$-regular with $k\ge3$, the ground state degeneracy grows exponentially in the number of vertices $\mathsf N$ \cite{alon90}.

Let us compare the Laplacian $\phi$-theory with a decoupled spin system: a finite dimensional spin at every site with trivial Hamiltonian. In both systems, the GSD grows exponentially in the number of vertices $\mathsf N$. How do we differentiate these systems? For simplicity, let us place both systems on a torus graph, $\Gamma = C_L \times C_L$. Then, in the Laplacian $\phi$-theory, the minimal number of generators of the discrete momentum symmetry group \eqref{lapphi-dismomsym} grows linearly in $L$ (see Section \ref{sec:2dlapphi}), and some of these generators have very large orders. On the other hand, in the decoupled spin system, there is a generator at each site, so the symmetry group has $L^2$ generators each with the same fixed order. In other words, the large GSD of Laplacian $\phi$-theory comes from the large orders of some of the generators of $\Jac(\Gamma)$ rather than the number of generators.

There is another way to derive the above ground state degeneracy. The discrete momentum and winding symmetries do not commute with each other: the shift \eqref{lapphi-momsym} of $n$ changes the discrete winding charge $\tilde{\mathcal Q}_a$ in \eqref{lapphi-modVill-discwind-charge} by $p_a \mod r_a$. This can be interpreted as a mixed 't Hooft anomaly between the discrete momentum and winding symmetries, and it leads to the ground state degeneracy \eqref{lapphi-gsd}. In fact, the entire Hilbert space is in a projective representation of   $\Jac(\Gamma)\times \Jac(\Gamma)$, so every state is $|\Jac(\Gamma)|$-fold degenerate.

\subsection{Spectrum}
Let us now determine the spectrum of the Laplacian $\phi$-theory \eqref{lapphi-modVill-action} by working with a continuous Lorentzian time while keeping the space discrete. To do this, we first take $L_\tau \rightarrow \infty$, and gauge fix $n_\tau(\hat \tau,i) = 0$, so that $n(\hat \tau,i) = n(i)$ and $k(\hat \tau,i) = k(i)$ are both time-independent. We then introduce the lattice spacing $a_\tau$ in the $\tau$-direction, and take the limit $a_\tau \rightarrow 0$ while keeping $\beta_0' = \beta_0 a_\tau$ and $\beta'=\beta/a_\tau$ fixed. Finally, we Wick rotate from Euclidean time $\tau$ to Lorentzian time $t$.

The equation of motion of $\phi$ is
\ie
\beta_0' \partial_0^2 \phi(t,i) + \beta' \Delta_L [\Delta_L \phi(t,i) - 2\pi n(i)] = 0~.
\fe
The general solution to this equation is
\ie\label{lapphi-eomsol}
&\phi(t,i) = f(i) + \phi_p(i) + \sum_{\lambda\ne0} \phi_\lambda(i) e^{i\omega_\lambda t}~,
\\
&n(i) = \frac{1}{2\pi} \Delta_L f(i) + p \delta_{i\mathsf N}~,
\fe
where $f(i)= 2\pi \sum_{a< \mathsf N} \frac{Q_{ia} p_a}{r_a} + c$ is a $U(1)$-valued discrete harmonic function given by \eqref{Laplace-sol-U1}, $\phi_p(i)$ is a real-valued solution to the equation
\ie
\Delta_L \phi_p(i) = 2\pi p \left(\delta_{i\mathsf N} - \frac{1}{\mathsf N}\right)~,\qquad p\in\mathbb Z~,
\fe
$\phi_\lambda(i)$ is a real-valued eigenfunction of the Laplacian operator with eigenvalue $\lambda$, i.e., $\Delta_L \phi_\lambda(i) = \lambda \phi_\lambda(i)$, and the dispersion relation for the ``plane wave modes'' is\footnote{We refer to them as the ``plane wave modes'' because when $\Gamma$ is a torus lattice in any dimension, the eigenvalues are related to the spatial momenta and the eigenfunctions form the usual Fourier basis.}
\ie
\omega_\lambda = \sqrt{\frac{\beta'}{\beta_0'}} \, \lambda~.
\fe
In other words, the plane wave spectrum is exactly the set of nonzero eigenvalues of the Laplacian operator. The smallest nonzero eigenvalue $\lambda_2$ of the Laplacian operator is known as the \emph{spectral gap} \cite{Chung:97}, or the \emph{algebraic connectivity} \cite{Fiedler1973} of the graph. When $\Gamma$ is a torus lattice in any dimension, $\lambda_2$ goes to zero with increasing number of sites. However, on a general graph, $\lambda_2$ could be finite even for large $\mathsf N$ \cite{NILLI1991207}. So the plane wave spectrum could be gapped on a general graph, while it is gapless on a torus lattice.

The zero mode $c$ of $f(i)$ in \eqref{lapphi-eomsol} is charged under the $U(1)$ momentum symmetry. After giving it a time dependence, its energy is lifted quantum mechanically to
\ie
E_\text{mom} = \frac{n^2}{2\beta_0'\mathsf N}~,
\fe
where $n \in\mathbb Z$ is the $U(1)$ momentum charge. See \cite{paper1} for a similar phenomenon, where a classical zero mode is lifted quantum mechanically, in another exotic model.

The rest of $f(i)$ in \eqref{lapphi-eomsol} is charged under the discrete winding symmetry $\Jac(\Gamma)$ with charges $\tilde{\mathcal Q}_a = p_a \mod r_a$ for $a<\mathsf N$. As we saw before, the discrete winding configurations have zero energy.

Finally, $\phi_p(i)$ in \eqref{lapphi-eomsol} is charged under the $U(1)$ winding symmetry with charge $\tilde{\mathcal Q} = p$. The energy of the winding configuration $\phi_p(i)$ is
\ie
E_\text{wind} = \frac{\beta'}{2} \sum_i [\Delta_L \phi_p(i) - 2\pi p\delta_{i\mathsf N}]^2 = \frac{(2\pi)^2\beta' p^2}{2\mathsf N}~.
\fe

\subsection{Robustness of GSD}
 
The action \eqref{lapphi-modVill-action} is said to be natural with respect to the global symmetry if all the relevant terms that are invariant under this symmetry are included in the action \cite{tHooft:1979rat}. (See \cite{paper1} for a recent discussion of naturalness and robustness). For example, a term that one can write on any graph is
\ie
-\sum_{\hat \tau,i} \cos[\phi(\hat \tau,i)]~.
\fe
However, this term is not invariant under the $U(1)$ momentum symmetry. So if we impose the $U(1)$ momentum symmetry, it is forbidden.

A more interesting term that one can write on any graph is the usual nearest-neighbor interaction\footnote{This is what one would write if one were studying the standard XY model on the graph $\Gamma$. In fact, let $\mathsf M$ denote the number of edges in $\Gamma$. Let us choose an orientation for each edge in $\Gamma$ arbitrarily so that we can talk about \emph{head} and \emph{tail} vertices of an edge. With respect to this orientation, the \emph{(oriented) incidence matrix} $B$ of $\Gamma$ is an $\mathsf N \times \mathsf M$ matrix given by $B_{i,e} = 1$ if $i$ is the head of $e$, $B_{i,e} = -1$ if $i$ is the tail of $e$, and $B_{i,e}=0$ otherwise. One can easily check that $B B^{T} = L$ independent of the choice of orientation. Therefore, \eqref{lapphi-edgeterm} is a ``lower order'' term with respect to the Laplacian term.}
\ie\label{lapphi-edgeterm}
-\sum_{\hat \tau, \langle i,j\rangle}\cos[ \phi(\hat \tau, i) - \phi(\hat \tau,j) ]~.
\fe
This term is clearly invariant under the $U(1)$ momentum symmetry, i.e., shifts by constants. What happens if we impose the full momentum symmetry $U(1)\times \Jac(\Gamma)$? When $\Gamma$ is a tree, the Jacobian group $\Jac(\Gamma)$ is trivial because $|\Jac(\Gamma)|=1$. In this case, \eqref{lapphi-edgeterm} is invariant, and so we should add this term to the action \eqref{lapphi-modVill-action}. This does not affect the ground state degeneracy because the latter is trivial anyway. On the other hand, when $\Gamma$ is not a tree, $\Jac(\Gamma)$ is nontrivial, which means the momentum symmetry includes non-constant shifts. So, \eqref{lapphi-edgeterm} is forbidden.

Now, consider the winding operator
\ie\label{lapphi-windterm}
-\sum_{\hat \tau,i}\cos [\tilde \phi(\hat \tau,i)]~.
\fe
This term is invariant under the $U(1)\times \Jac(\Gamma)$ momentum symmetry, so if we impose only the momentum symmetry, we should add it to the action \eqref{lapphi-modVill-action}. Indeed, \eqref{lapphi-windterm} is relevant because it breaks the $U(1)$ winding symmetry and lifts the ground state degeneracy. In other words, if we impose only the momentum symmetry, the winding symmetry is not robust.

It is also natural to impose both momentum and winding symmetries. In this case, \eqref{lapphi-windterm} is forbidden and the ground state degeneracy cannot be lifted.

\subsection{Examples}

In this section, we discuss the Laplacian $\phi$-theory on 1d and 2d spatial tori.

\subsubsection{1+1d dipole $\phi$-theory}\label{sec:1ddipphi}

\begin{figure}
\begin{center}
\raisebox{0.08\height}{\includegraphics[scale=0.24]{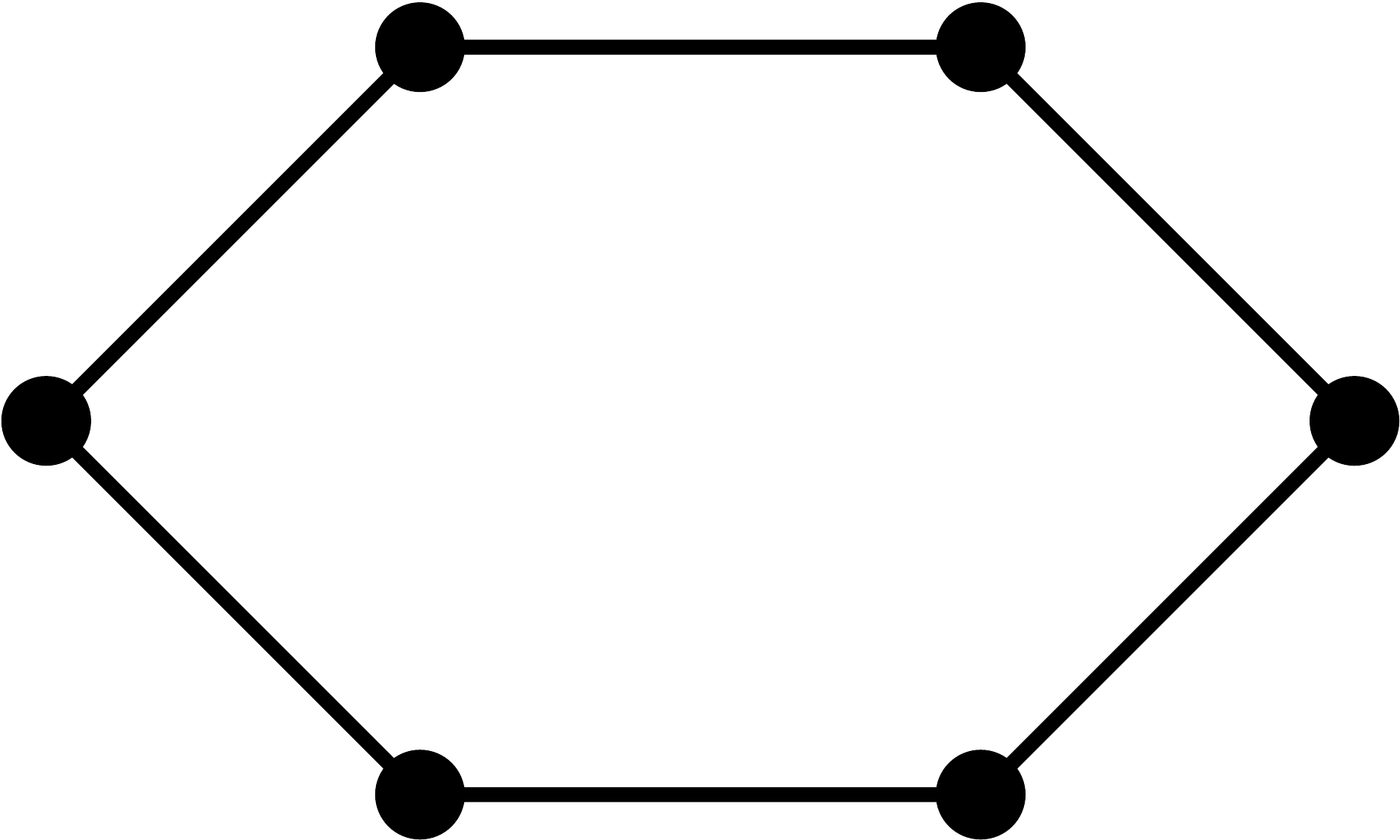}} ~~~~ \includegraphics[scale=0.12]{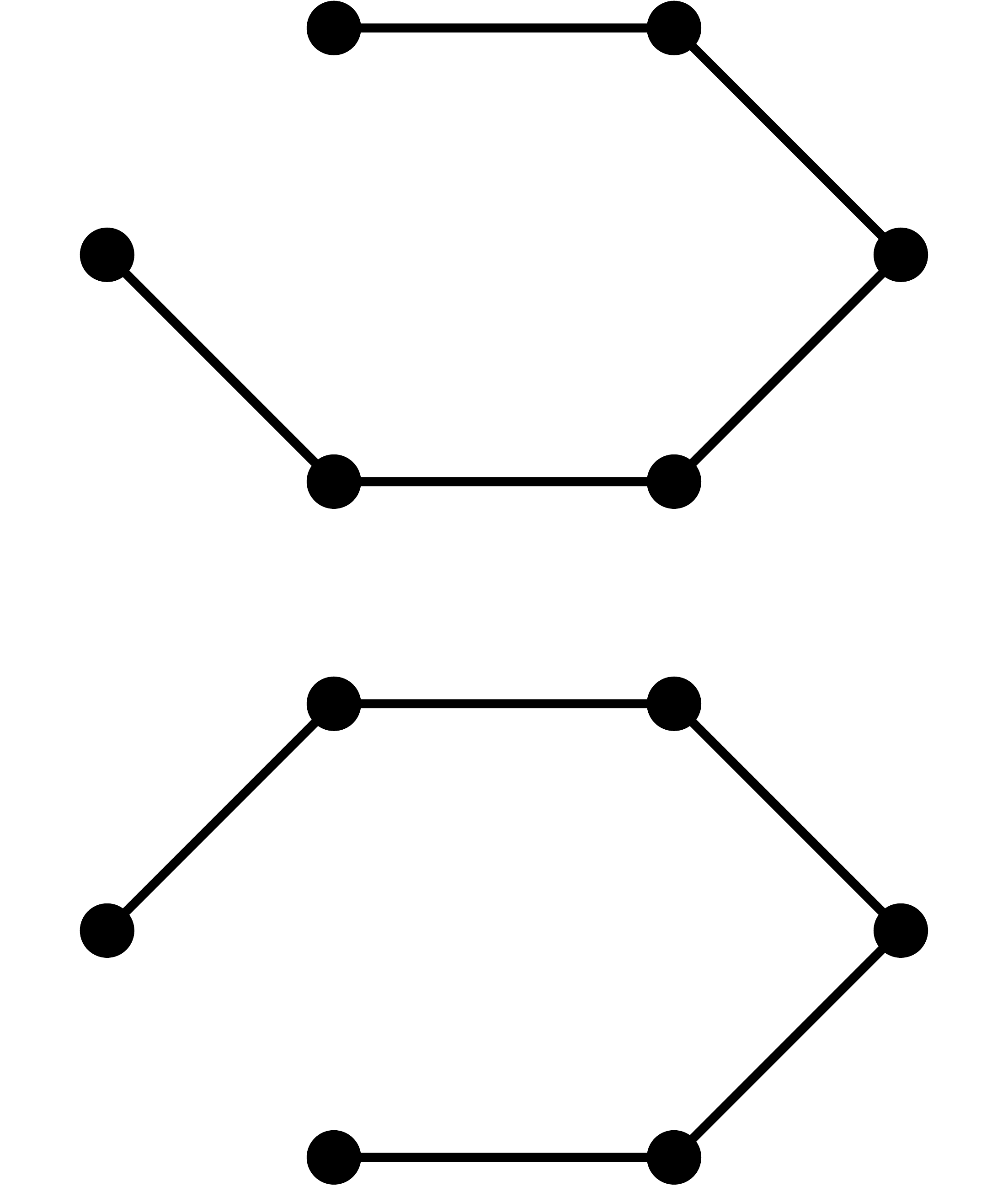}~~\includegraphics[scale=0.12]{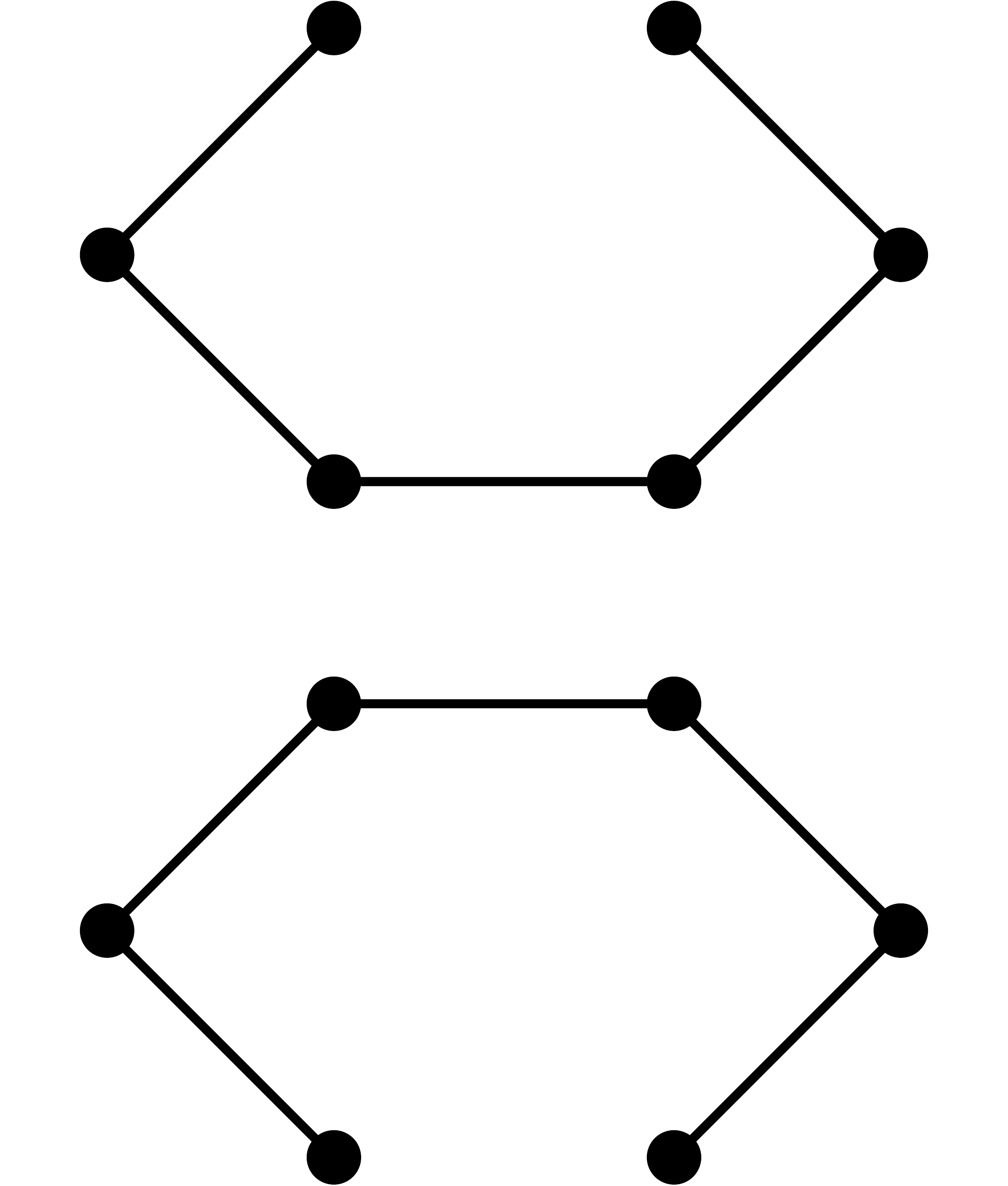}~~\includegraphics[scale=0.12]{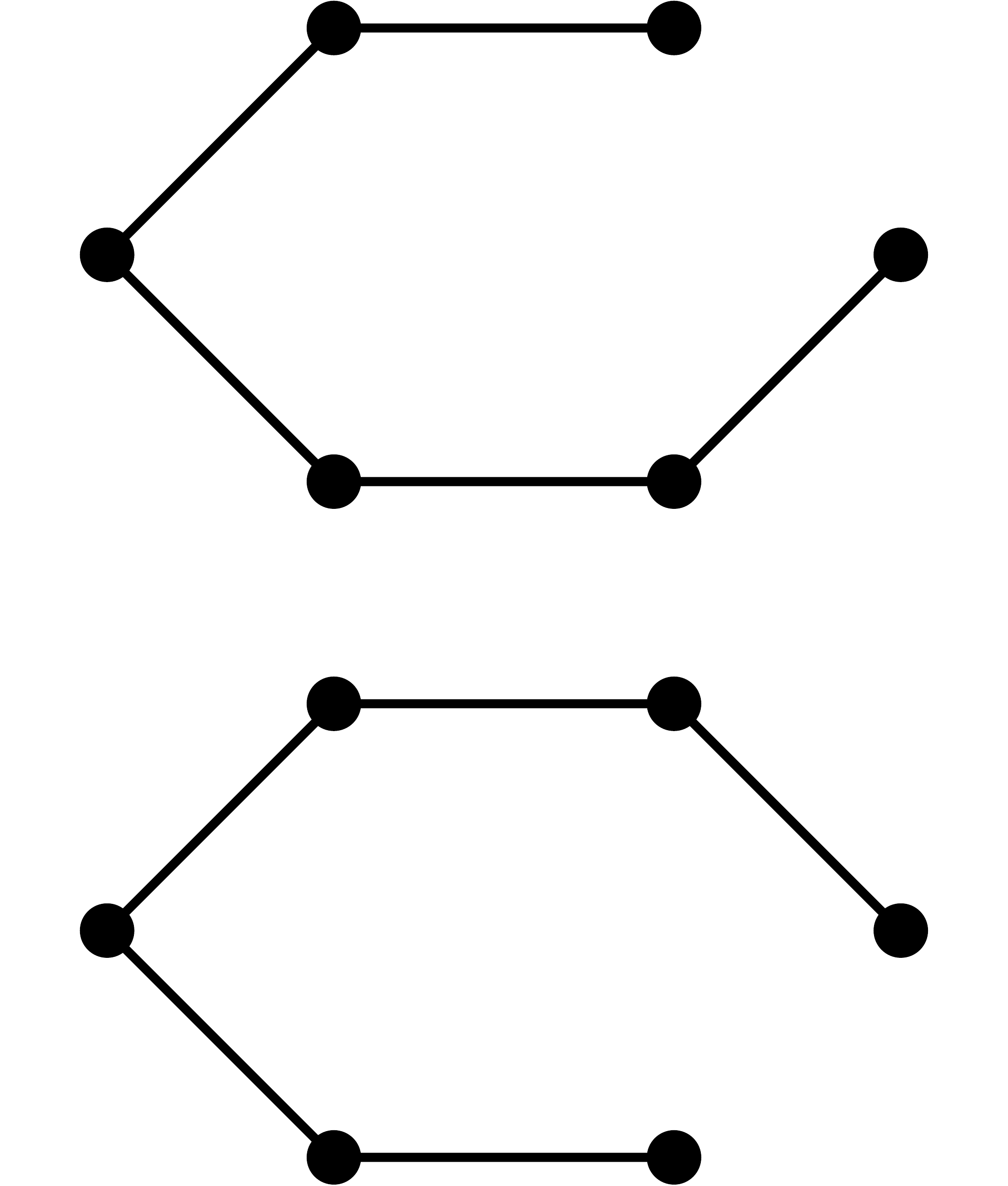}
\\
(a) ~~~~~~~~~~~~~~~~~~~~~~~~~~~~~~~~~~~~~~~~~~~~~~~~~(b)~~~~~~~~~~~~~~~~
\end{center}
\caption{(a) The cycle graph $C_6$ on $L_x = 6$ vertices. (b) The six spanning trees of $C_6$ obtained by removing one of the six edges.}\label{fig:cyclegraph}
\end{figure}

Let $\Gamma$ be a \emph{cycle graph} $C_{L_x}$ (see Figure \ref{fig:cyclegraph}), i.e., $\Gamma = C_{L_x} = \mathbb Z_{L_x}$, where $L_x$ is the number of sites in the cycle. The operator $\Delta_L$ associated with the Laplacian matrix of $\Gamma$ is the same as the standard Laplacian operator $\Delta_x^2$ in the $x$-direction.\footnote{Actually, it is common to define $\Delta_x^2 f(\hat x) = f(\hat x+1) - 2f(\hat x) + f(\hat x-1)$, so $\Delta_L = - \Delta_x^2$. We will ignore this discrepancy in sign since it does not affect the rest of the discussion.} The action \eqref{lapphi-modVill-action} becomes
\ie
S &= \frac{\beta_0}{2} \sum_{\tau\text{-link}} \left(\Delta_\tau \phi- 2\pi n_\tau \right)^2 + \frac{\beta}{2} \sum_\text{site} \left( \Delta_x^2 \phi - 2\pi n_{xx} \right)^2
\\
& \qquad + i\sum_{\tau\text{-link}} \tilde \phi \left( \Delta_\tau n_{xx} - \Delta_x^2 n_\tau \right)~,
\fe
after replacing $n$ with $n_{xx}$. This is the modified Villain action of the 1+1d dipole $\phi$-theory \cite{Gorantla:2022eem}, which is the lattice regularization of the 1+1d compact Lifshitz scalar field theory. 

The invariant factors of $L$ are 
\ie
r_a = \begin{cases}
	1~,& 1\le a < L_x -1~,\\
	L_x~,& a = L_x-1~,\\
	0~,& a = L_x~.
\end{cases}
\fe
It follows that the Jacobian group for a cycle graph $C_{L_x}$ is
\ie\label{1ddipphi-Jac}
\Jac(C_{L_x}) = \mathbb Z_{L_x}~.
\fe
\eqref{1ddipphi-Jac} is a demonstration of the fact that the large GSD comes from the large orders of some of the genrators of $\Jac(\Gamma)$ rather than the number of generators. Physically, \eqref{1ddipphi-Jac} means that the discrete momentum and winding symmetries of the 1+1d dipole $\phi$-theory are $\mathbb Z_{L_x}$.  

A spanning tree of a cycle graph $C_{L_x}$ is obtained by removing any one of the $L_x$ edges. See Figure \ref{fig:cyclegraph}. Therefore, there are exactly $L_x$ spanning trees of a cycle graph $C_{L_x}$. In other words, the complexity of $C_{L_x}$ is simply $L_x$, which equals the ground state degeneracy of the 1+1d dipole $\phi$-theory. This is in agreement with the analysis of \cite{Gorantla:2022eem}.
As both   the discrete global symmetry (which is $\Jac(C_{L_x})$) and the ground state degeneracy (which is $|\Jac(C_{L_x})|$) grow as we increase the number of lattice sites $L_x$, it is clear that  this 1+1d model does not have an unambiguous continuum limit $L_x\to\infty$.

Incidentally, using \eqref{graph-matrixtree} to compute the ground state degeneracy in terms of the eigenvalues of $\Delta_x^2$ leads to the following identity:
\ie\label{1ddipphi-identity}
L_x = \frac{1}{L_x}\prod_{k_x = 1}^{L_x - 1} 4\sin^2\left(\frac{\pi k_x}{L_x}\right)~.
\fe

\subsubsection{2+1d Laplacian $\phi$-theory}\label{sec:2dlapphi}
Let $\Gamma$ be a \emph{torus graph}, i.e., $\Gamma = C_{L_x} \times C_{L_y} = \mathbb Z_{L_x}\times \mathbb Z_{L_y}$, where $L_i$ is the number of sites in the $i$ direction. The operator $\Delta_L$ associated with the Laplacian matrix of $\Gamma$ is the same as the standard Laplacian operator $\Delta_x^2 + \Delta_y^2$ in the $xy$-plane. The action \eqref{lapphi-modVill-action} becomes
\ie
S &= \frac{\beta_0}{2} \sum_{\tau\text{-link}} \left(\Delta_\tau \phi- 2\pi n_\tau \right)^2 + \frac{\beta}{2} \sum_\text{site} \left[ (\Delta_x^2+\Delta_y^2) \phi - 2\pi n \right]^2
\\
& \qquad + i\sum_{\tau\text{-link}} \tilde \phi \left[ \Delta_\tau n - (\Delta_x^2 + \Delta_y^2) n_\tau \right]~.
\fe
We refer to this as the 2+1d Laplacian $\phi$-theory. This is a natural lattice regularization of the 2+1d compact Lifshitz scalar field theory that can be defined on a general spatial graph $\Gamma$.

The discrete momentum and winding symmetries are determined by the Jacobian group of the torus graph. The latter is quite complicated and not known in closed form in general as a function of $L_x,L_y$. Below we record a few examples for small values of $L_x = L_y$:
\ie
&\Jac(C_2 \times C_2) = \mathbb Z_2 \times \mathbb Z_2 \times \mathbb Z_8~,
\\
&\Jac(C_3 \times C_3) = \mathbb Z_6 \times \mathbb Z_6 \times \mathbb Z_{18} \times \mathbb Z_{18}~,
\\
&\Jac(C_4 \times C_4) = \mathbb Z_2 \times \mathbb Z_2 \times \mathbb Z_8 \times \mathbb Z_{24} \times \mathbb Z_{24} \times \mathbb Z_{24} \times \mathbb Z_{96}~.
\fe

 The minimal number of generators of $\Jac(C_{L_x} \times C_{L_y})$ is at most the number of nontrivial spatial integer gauge fields, $n$'s, after gauge fixing. One can gauge fix the $n$'s to be zero everywhere except along $\hat x = 0,1$, or along $\hat y = 0,1$. Therefore, after gauge fixing, the number of nontrivial $n$'s is $\min(2L_x,2L_y)$. In other words, on a square torus graph $C_L \times C_L$, the minimal number of generators of $\Jac(C_L \times C_L)$ grows at most linearly in $L$. Once again, this shows that the large GSD is due to the large orders of some of the generators of $\Jac(\Gamma)$ rather than the number of generators.

The ground state degeneracy is given by the order of the Jacobian group, $\text{GSD}=|\Jac(C_{L_x}\times C_{L_y})|$. While there is no closed form formula for the Jacobian group itself, the order of this group can be expressed in terms of the eigenvalues of the discrete Laplacian on the torus graph (see \eqref{graph-matrixtree}). We therefore obtain
\ie
\text{GSD} &= \frac{1}{L_x L_y} \prod_{0\le k_i <L_i\atop(k_x,k_y)\ne(0,0)} \left[4\sin^2\left( \frac{\pi k_x}{L_x} \right)+4\sin^2\left( \frac{\pi k_y}{L_y} \right)\right]
\\
&= L_x L_y \prod_{k_i = 1}^{L_i-1} \left[4\sin^2\left( \frac{\pi k_x}{L_x} \right)+4\sin^2\left( \frac{\pi k_y}{L_y} \right)\right]~,
\fe
where we used the identity \eqref{1ddipphi-identity} in the second line. Since the discrete global symmetry $\Jac(C_{L_x}\times C_{L_y})$ and the GSD depend sensitively on the number theoretic properties of the lattice sites $L_x,L_y$, this regularization of the 2+1d compact Lifshitz theory does not have an unambiguous continuum limit $L_x,L_y\to\infty$.

Let us discuss the asymptotic behavior of the GSD on a torus. The GSD grows asymptotically as \cite{Wu_1977}
\ie
\log \text{GSD} \approx \frac{L_x L_y}{\pi^2} \int_0^\pi \int_0^\pi dp_x dp_y ~ \log \left[ 4 \sin^2(p_x) + 4 \sin^2(p_y) \right] = \frac{4 G }{\pi}L_x L_y~,
\fe
where $G\approx 0.916$ is the Catalan constant. This is consistent with the intuition that the number of spanning trees of the torus graph $C_{L_x}\times C_{L_y}$, which is also equal to GSD, grows with $L_x$ and $L_y$. Indeed, the result of \cite{alon90} implies the ground state degeneracy of the 2+1d Laplacian $\phi$-theory grows exponentially in $L_x L_y$.\footnote{Here we have used the fact that the torus graph $\Gamma = C_{L_x}\times C_{L_y}$ is $4$-regular in applying the result of \cite{alon90}. In contrast, a cycle graph is only 2-regular, and indeed the GSD of the 1+1d dipole $\phi$-theory grows linearly in the number of vertices.}

A closely related matter theory is the 2+1d dipole $\phi$-theory, which is another possible regularization of the 2+1d compact Lifshitz field theory. While the 1+1d dipole $\phi$-theory is the same as the 1+1d Laplacian $\phi$-theory, their 2+1d versions are very different. In an upcoming paper \cite{Gorantla:2022ssr}, we compare the 2+1d Laplacian and dipole $\phi$-theories, and discuss their relation to the 2+1d compact Lifshitz theory.

\section{$U(1)$ Laplacian gauge theory and fractons on a graph}\label{sec:lapA}
We can gauge the momentum symmetry of the Laplacian $\phi$-theory by coupling it to the gauge fields $(\mathcal A_\tau,\mathcal A;m_\tau)$. Here $\mathcal{A}_\tau$ and $\mathcal{A}$ are real-valued fields living on the $\tau$-links and the sites, respectively, and $m_\tau$ is an integer-valued gauge field living on the $\tau$-link. Their gauge transformations are
\ie\label{lapA-gauge}
&\mathcal A_\tau \sim \mathcal A_\tau + \Delta_\tau \alpha + 2\pi q_\tau~,
\\
&\mathcal A \sim \mathcal A + \Delta_L \alpha + 2\pi q~,
\\
&m_\tau \sim m_\tau + \Delta_\tau q - \Delta_L q_\tau~,
\fe
where $\alpha$ is a real-valued gauge parameter and $q_\tau,q$ are integer-valued gauge parameters.

We can leave out the matter fields and study the pure gauge theory of $(\mathcal A_\tau,\mathcal A;m_\tau)$. It is described by the following Villain action
\ie\label{lapA-modVill-action}
S = \frac{\gamma}{2} \sum_{\hat \tau,i} \mathcal E^2 + \frac{i\theta}{2\pi} \sum_{\hat \tau,i} \mathcal E~,
\fe
where $\mathcal E = \Delta_\tau \mathcal A - \Delta_L \mathcal A_\tau - 2\pi m_\tau$ is the gauge-invariant electric field of $(\mathcal A_\tau,\mathcal A;m_\tau)$. The $\theta$-angle is $2\pi$-periodic, i.e.,  $\theta \sim \theta + 2\pi$, because $\sum_{\hat \tau,i} \mathcal E = -2\pi \sum_{\hat \tau,i} m_\tau \in 2\pi \mathbb{Z}$. See \cite{Gorantla:2021svj} for similar Villain formulation of various standard and exotic  $U(1)$ gauge theories. 

The theory \eqref{lapA-modVill-action} can be viewed as an extension of the 1+1d rank-2 $U(1)$ gauge theory of \cite{Gorantla:2022eem} with $\Delta_x^2$ on the 1d spatial lattice replaced by the discrete Laplacian operator $\Delta_L$ on the graph $\Gamma$.

\subsection{Global symmetry}
Here we discuss the global symmetry of the $U(1)$ Laplacian gauge theory. There are two kinds of global symmetries that we should distinguish in gauge theory. The first kind is the \textit{space-like} global symmetry, which acts on operators and states in the Hilbert space. The second kind is the \textit{time-like} global symmetry, which acts on defects extended in the (Euclidean) time direction. In an ordinary, relativistic gauge theory, both the space-like and time-like global symmetries are parts of the one-form global symmetry \cite{Gaiotto:2014kfa}. In contrast, the two global symmetries can be drastically different in non-relativistic systems. For example, even the groups for them can be different. We refer the readers to \cite{Gorantla:2022eem} for comprehensive analyses of time-like global symmetries in various standard and exotic models.

This theory has an electric symmetry that shifts
\ie
(\mathcal A_\tau, \mathcal A; m_\tau) \rightarrow (\mathcal A_\tau, \mathcal A; m_\tau) + (\lambda_\tau, \lambda; p_\tau)~,
\fe
where $(\lambda_\tau, \lambda; p_\tau)$ is a flat $U(1)$ gauge field, i.e., 
\ie
\Delta_\tau \lambda - \Delta_L \lambda_\tau = 2\pi p_\tau\,.
\fe

The shift $(\lambda_\tau, \lambda ; p_\tau)$ is subject to the gauge transformation  in \eqref{lapA-gauge}. Below we will use this freedom of gauge transformation to gauge-fix the shift in a particular form. Using the integer gauge parameter $q$, we can set $p_\tau = 0$ everywhere except at $\hat \tau = 0$. Similarly, we can use $\alpha$ to set $\lambda_\tau = 0$ everywhere except at $\hat \tau = 0$. Since $\sum_i p_\tau(\hat \tau,i) = 0$ by flatness, by the analysis in Appendix \ref{app:lap-poisson} around \eqref{gaugefix-Z}, we can use $q_\tau$ to set 
\ie
p_\tau(\hat \tau,i) = -\delta_{\hat \tau,0} \sum_{a<\mathsf N} (P^{-1})_{ia} p_{\tau a}~,
\fe
where $p_{\tau a} = 0,\ldots, r_a-1$. Now, the remaining gauge symmetry is time-independent $\alpha(i)$ and $q(i)$, and $q_\tau(\hat \tau,i) = \delta_{\hat \tau,0}\bar q_\tau$, where $\bar q_\tau$ is an integer. By flatness, we have $\Delta_\tau \lambda = 0$. By the analysis in Appendix \ref{app:lap-poisson} around \eqref{gaugefix-U1}, using $\alpha(i)$ and $q(i)$, we can set 
\ie\label{lambda}
\lambda(\hat \tau,i) = \frac{c}{\mathsf N}~,
\fe
where $c\sim c+2\pi$. 

Since $\Delta_\tau \lambda = 0$, we have $\Delta_L \lambda_\tau + 2\pi p_\tau = 0$ at $\hat \tau = 0$. Using \eqref{Laplace-sol-U1}, the solution for $\lambda_\tau$ is
\ie\label{lambdatau}
\lambda_\tau(\hat \tau,i) = \delta_{\hat \tau,0} \left[ c_\tau + 2\pi \sum_{a< \mathsf N} \frac{Q_{ia} p_{\tau a}}{r_a} \right]~,
\fe
where $c_\tau \sim c_\tau + 2\pi$.

The parameter $c$ generates a $U(1)$ electric global space-like symmetry that shifts $\mathcal A \rightarrow \mathcal A + \frac{c}{\mathsf N}$. The operator charged under this symmetry is
\ie
\exp\left[ i\sum_j \mathcal A(\hat \tau,j) \right]~.
\fe
There are no other gauge invariant operators other than those mentioned so far. Relatedly, there is no discrete electric global symmetry.

The parameters $c_\tau$ and $p_{\tau a}$'s generate the electric global time-like symmetry that shifts
\ie\label{lapA-timelike}
&\mathcal A_\tau (\hat\tau, i)\rightarrow \mathcal A_\tau (\hat\tau, i)+ \delta_{\hat \tau,0} \left[ c_\tau + 2\pi \sum_{a< \mathsf N} \frac{Q_{ia} p_{\tau a}}{r_a} \right]~,
\\
&m_\tau(\hat\tau, i) \rightarrow m_\tau(\hat\tau, i) - \delta_{\hat \tau,0} \sum_{a<\mathsf N} (P^{-1})_{ia} p_{\tau a}~.
\fe
Comparing with \eqref{Laplace-sol-U1}, we see that a general time-like global symmetry transformation is labeled by  a $U(1)$-valued  discrete harmonic function, which forms the group  $U(1) \times \Jac(\Gamma)$. This shift does not act on operators. Rather, it acts on defects that extend in the time direction. More specifically, a  charged defect  at site $i$ is 
\ie\label{lapA-defect}
\exp\left[ i\sum_{\hat \tau} \mathcal A_\tau(\hat \tau,i) \right]~.
\fe

\subsection{Spectrum}\label{sec:U1spectrum}

Let us now determine the spectrum of the $U(1)$ Laplacian gauge theory \eqref{lapA-modVill-action} by working with continuous Lorentzian time while keeping the space discrete. To do this, we first take $L_\tau \rightarrow \infty$, and gauge fix $m_\tau(\hat \tau,i) = 0$. We then introduce the lattice spacing $a_\tau$ in the $\tau$-direction, and take the limit $a_\tau \rightarrow 0$ while keeping $\gamma' = \gamma a_\tau$ fixed. We also define a scaled temporal gauge field $A_\tau = \frac{1}{a_\tau} \mathcal A_\tau$ and a scaled electric field $E = \partial_\tau \mathcal A - \Delta_L A_\tau$ while taking this limit. Finally, we Wick rotate from Euclidean time $\tau$ to Lorentzian time $t$.

The Gauss law (i.e., the equation of motion of $A_0$) gives
\ie
\Delta_L E(t,i) = 0 \implies E(t,i) = E(t)~.
\fe
In the temporal gauge $A_0(t,i) = 0$, up to a time-independent gauge transformation, this equation is solved by
\ie
\mathcal A(t,i) = \frac{c(t)}{\mathsf N}~,
\fe
where $c(t)$ is circle-valued, i.e., $c(t) \sim c(t) + 2\pi$. The effective action for $c(t)$ is
\ie
S = \int dt~ \left[\frac{\gamma'}{2\mathsf N} \dot c(t)^2 - \frac{\theta}{2\pi} \dot c(t) \right]~.
\fe
The Hamiltonian is
\ie
H = \frac{\mathsf N}{2\gamma'}\left(\Pi + \frac{\theta}{2\pi} \right)^2~,
\fe
where $\Pi$ is the conjugate momentum of $c(t)$, and $\Pi \in \mathbb Z$ because of the periodicity of $c(t)$.  
 For $\theta\neq \pi$, the ground state is non-degenerate, while for $\theta=\pi$ there are two degenerate ground states. 

We now discuss the robustness of the theory. All the local operators in the theory are made of the gauge-invariant electric field $E$. Adding them to the Lagrangian does not change the qualitative behavior of the theory. Hence, we conclude that the theory is robust.

\subsection{Mobility of defects: fractons}

The pure $U(1)$ gauge theory has defects \eqref{lapA-defect} describing the world lines of infinitely massive particles. 
The $U(1)$ time-like symmetry charges of these defects can be interpreted as the gauge charges of the massive particles. 
More generally, a static configuration of particles carrying gauge charge $q(i)$ at site $i$  is represented by the following defect: 
\ie\label{divisorcorrespondence}
\exp\left[
i\sum_{\hat \tau} \sum_i q(i) \mathcal A_\tau(\hat\tau,i)\right]~,\qquad q(i)\in \mathbb{Z}~.
\fe
A ``move" at time $\hat\tau_0$ on a configuration is implemented by applying products of operators $\exp[i  {\cal A}(\hat\tau_0 ,i)]$ at different sites. 
A configuration of particles can ``move" to another configuration if and only if there is a gauge-invariant defect that connects the two.

Since the discrete $\text{Jac}(\Gamma)$ time-like global symmetries depend on the sites, they constrain the possible shapes of defects, and therefore the mobility of the particles. Moreover, they lead to superselection sectors of defects distinguished by the time-like symmetry charges.  
See \cite{Gorantla:2022eem} for more discussions on time-like global symmetries. 

From \eqref{lapA-timelike}, we see that the discrete time-like symmetry charges of a static defect \eqref{lapA-defect} at site $i$ are given by $Q_{ia}$ mod $r_a$ with $a=1,2,\cdots, \mathsf N-1$. The defect \eqref{lapA-defect} can hop from site $i$ to site $i'$ if and only if the time-like charges of the defects at these two positions are the same, i.e., 
\ie\label{lapA-mobilecond}
Q_{i'a} = Q_{ia} \mod r_a~,\qquad a=1,\ldots,\mathsf N-1~.
\fe
Then, the defect that ``hops'' a particle from $i$ to $i'$ at time $\hat \tau=\hat \tau_0$ is
\ie\label{lapA-mobiledefect}
\exp\left[ i \sum_{\hat \tau<\hat \tau_0} \mathcal A_\tau(\hat \tau,i) \right]\exp\left[ i \sum_{a<\mathsf N,j} \left(\frac{Q_{i'a} - Q_{ia}}{r_a}\right) P_{aj} \mathcal A(\hat \tau_0,j) \right]\exp\left[ i \sum_{\hat \tau\ge\hat \tau_0} \mathcal A_\tau(\hat \tau,i') \right]~.
\fe

We are now ready to phrase the mobility of the probe particles in terms of a graph-theoretic statement. 
The condition for mobility \eqref{lapA-mobilecond} is equivalent to the property that \textit{all} $U(1)$-valued discrete harmonic functions \eqref{Laplace-sol-U1} take the same value at $i$ and $i'$. If the condition \eqref{lapA-mobilecond} is not satisfied, i.e., if there is a $U(1)$-valued discrete harmonic function that takes different values at $i$ and $i'$, then the particle cannot move from $i$ to $i'$. If this is the case for all $i,i'\in\Gamma$, then the particle is a \emph{fracton}.

Using the Abel-Jacobi map $S_{i_0}$, we can completely characterize the mobility of a particle by general properties of the graph $\Gamma$.  (See Section \ref{sec:graph} for the definition and properties of the Abel-Jacobi map $S_{i_0}$, where $i_0$ is some fixed vertex of the graph.)

\begin{itemize}
\item
If there is a unique path from $i$ to $i'$, then $S_{i_0}(i) = S_{i_0}(i')$. By the universal property of $S_{i_0}$, for any $U(1)$-valued discrete harmonic function $f$ that vanishes at $i_0$, there is a unique group homomorphism $\psi_f:\Jac(\Gamma)\rightarrow U(1)$ such that $f=\psi_f\circ S_{i_0}$. In particular, $f(i) = f(i')$ for any $U(1)$-valued discrete harmonic function $f\in\mathcal H(\Gamma,U(1))$, so there is no selection rule imposed by the time-like symmetry for a particle to move from $i$ to $i'$. Indeed, Figure \ref{fig:fracton} illustrates how a particle can move from $i$ to $i'$ in this case.

\begin{figure}[t!]
\begin{center}
\begin{tikzcd}[row sep=1cm,column sep=3.2cm]
\includegraphics[scale=0.17]{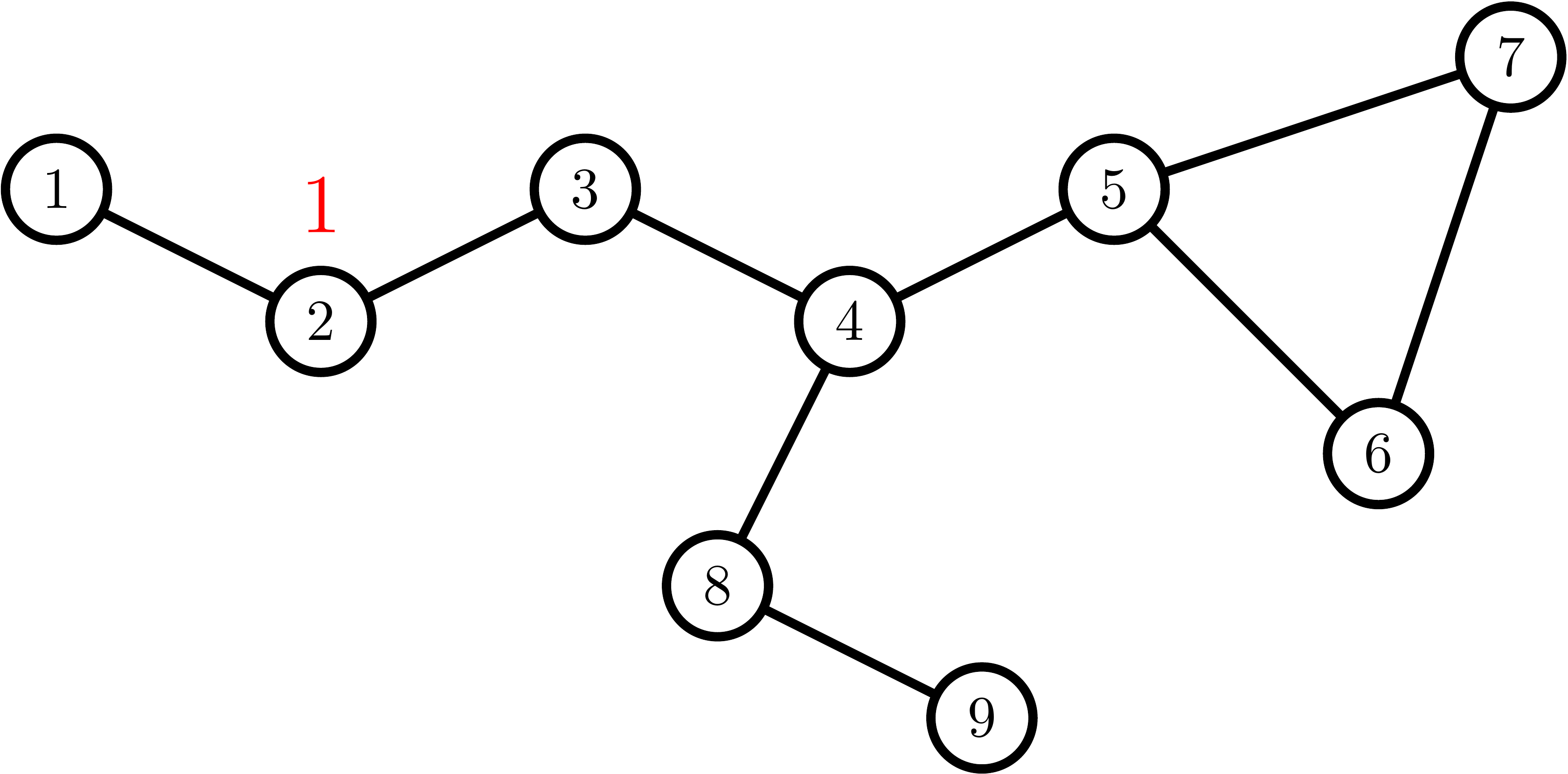}\arrow{d}[swap]{\exp[i\mathcal A(\hat\tau_0,3)]} & \includegraphics[scale=0.17]{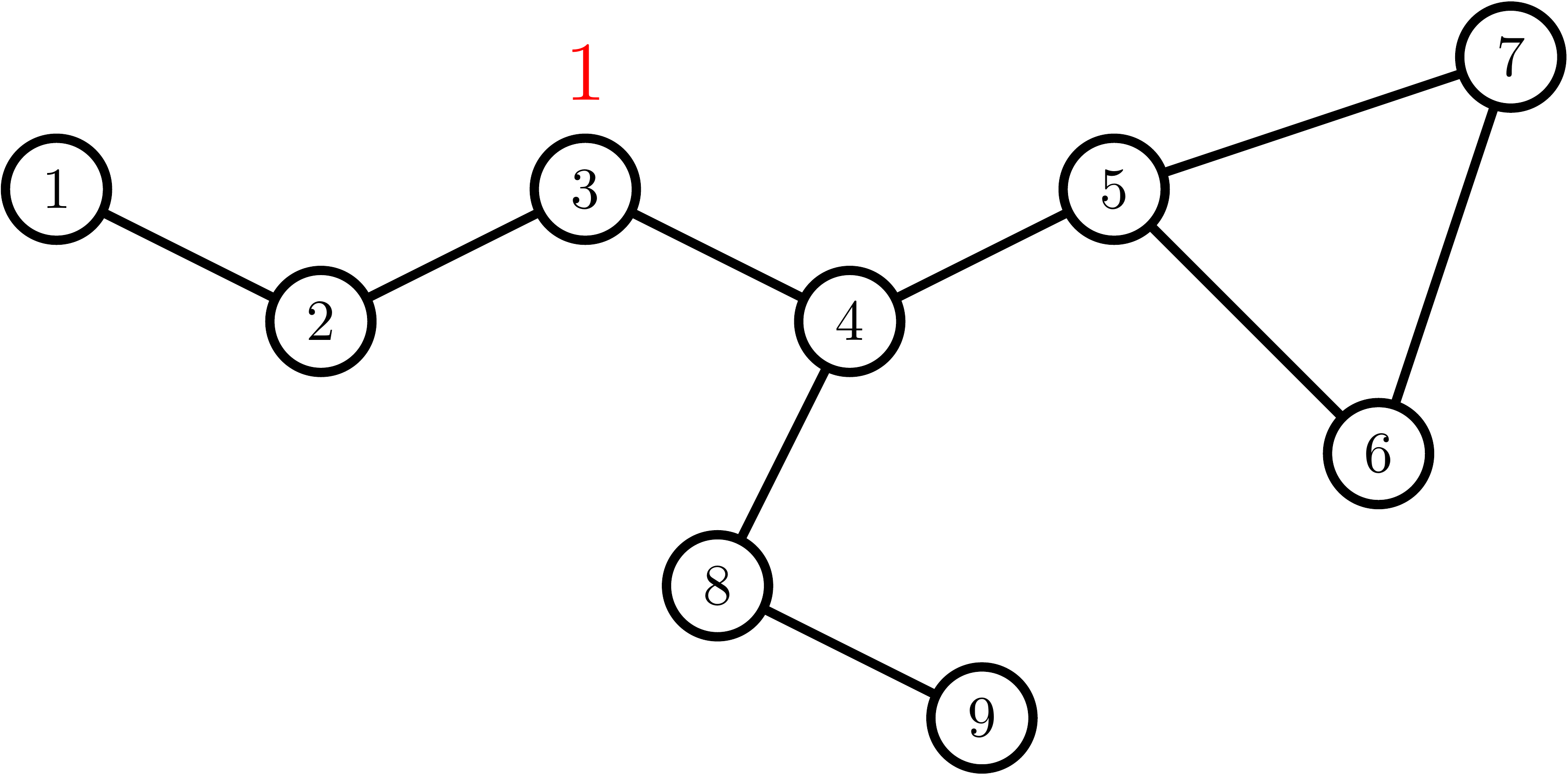}
\\
\includegraphics[scale=0.17]{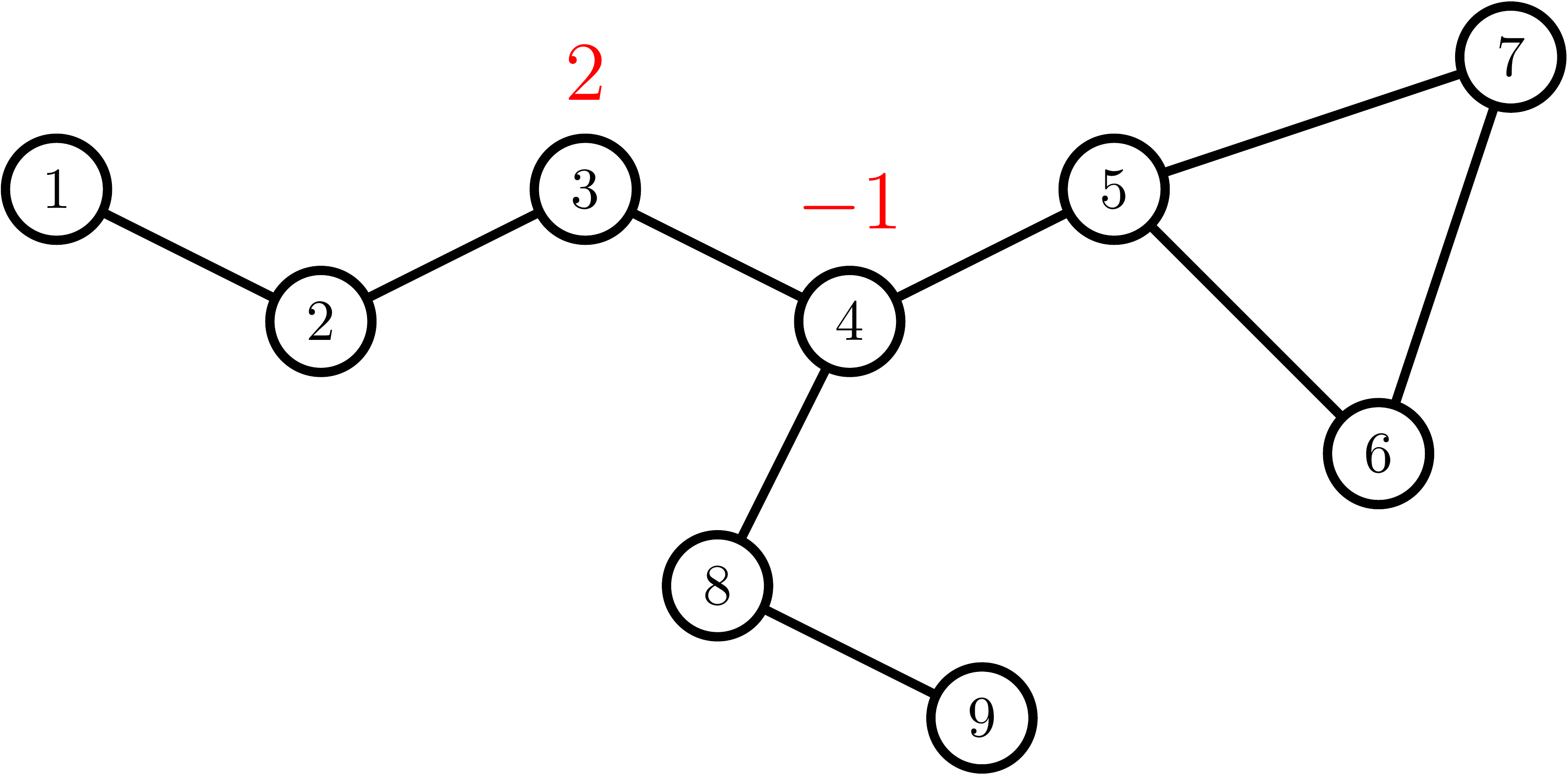}\arrow{d}[swap]{\exp[i\mathcal A(\hat\tau_0,4)]} & \includegraphics[scale=0.17]{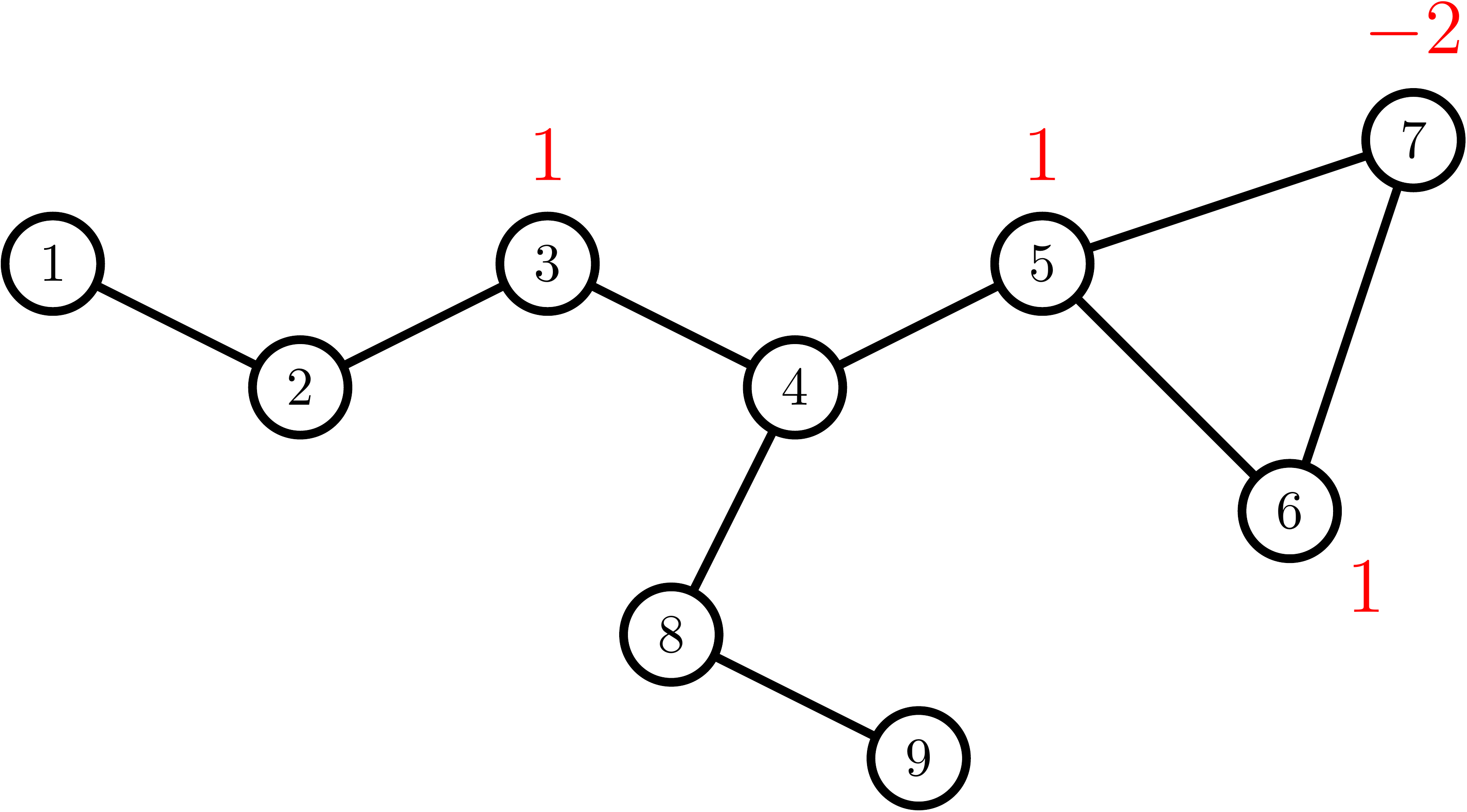}\arrow{u}[swap]{\exp[i\mathcal A(\hat\tau_0,7)]}
\\
\raisebox{-0.4\height}{\includegraphics[scale=0.17]{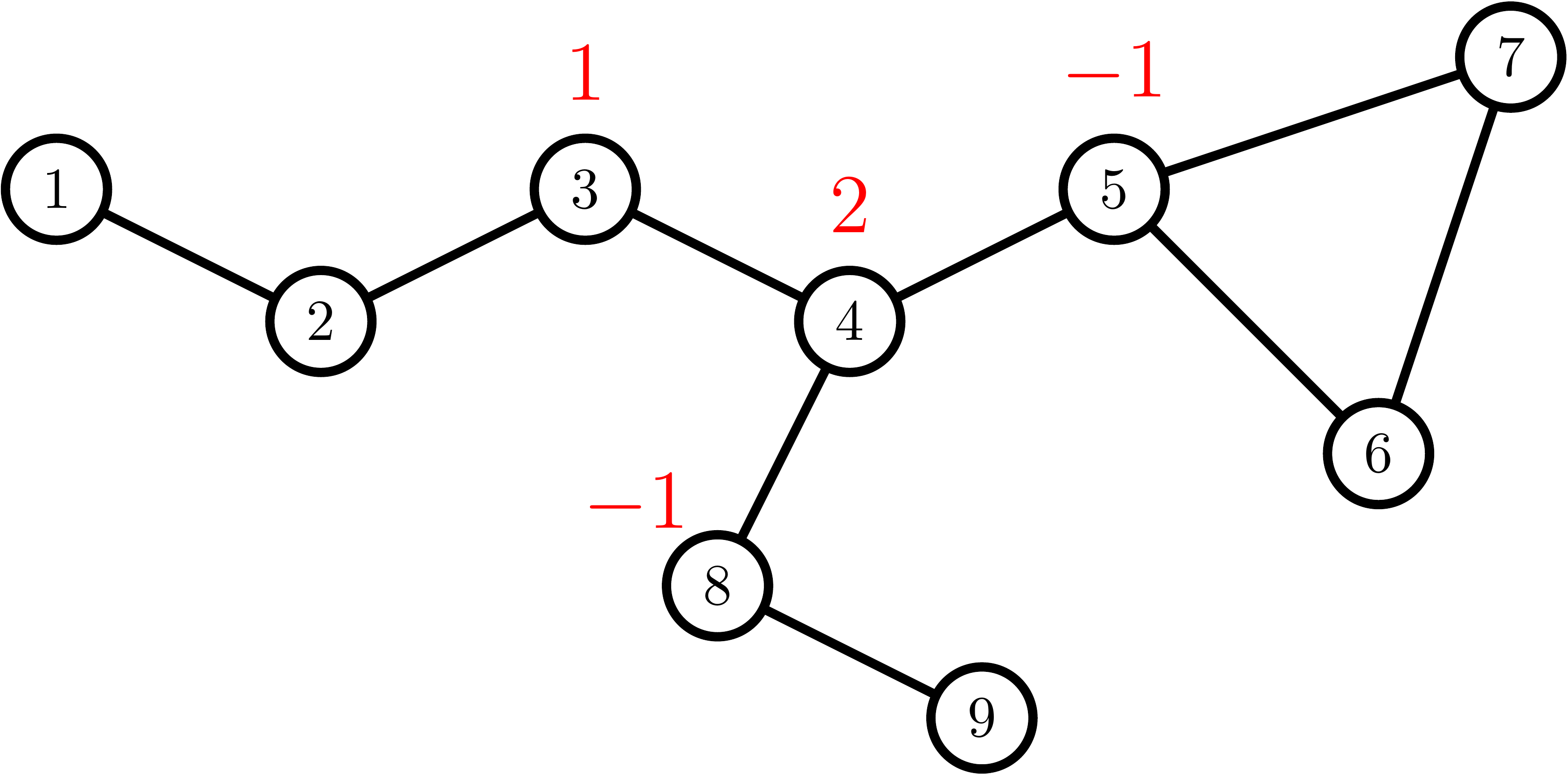}}\arrow{r}{\exp[i\mathcal A(\hat\tau_0,5)+i\mathcal A(\hat\tau_0,8)]} & \raisebox{-0.4\height}{\includegraphics[scale=0.17]{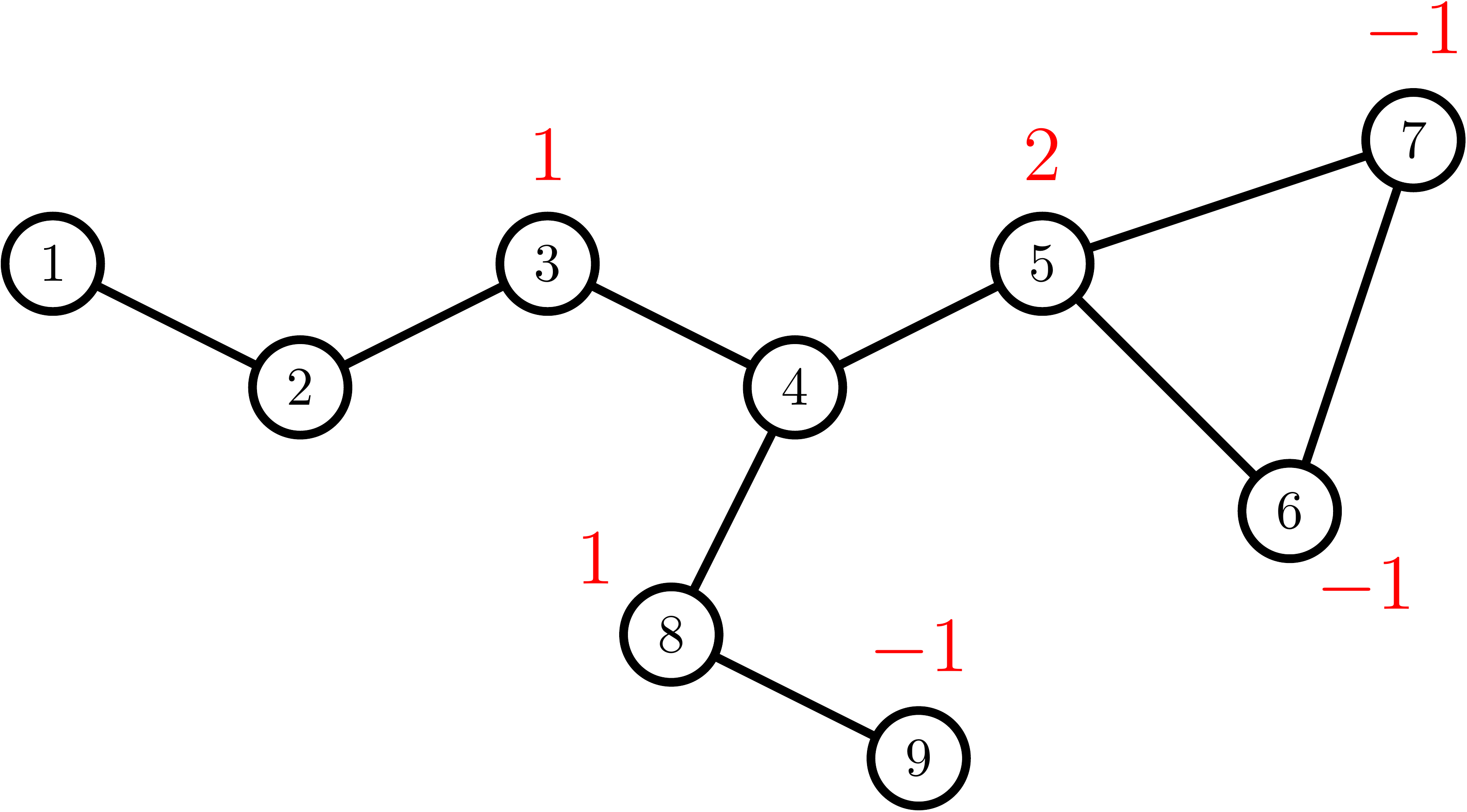}}\arrow{u}[swap]{\exp[i\mathcal A(\hat\tau_0,6)+i\mathcal A(\hat\tau_0,9)]}
\end{tikzcd}
\end{center}
\caption{Motion of a particle between the vertices $2$ and $3$ that are connected by a unique path. In the graph, the numbers in black label the vertices, and the integers in red denote the charges of the particles (a zero charge is omitted). The operator $\exp[i\mathcal A(\hat\tau_0,i)]$ creates a particle of charge $d_i$ at vertex $i$ (where $d_i$ is the degree of the vertex $i$) and particles of charge $-1$ at the neighbors of $i$ at a fixed time $\hat\tau_0$. The above sequence of operators moves the particle from vertex $2$ to vertex $3$. In the correspondence between fractons and divisors, the integers in red represent values of the divisor (a zero value is omitted), and the operator $\exp[i\mathcal A(\hat\tau_0,i)]$ changes the divisor by a principal divisor.}\label{fig:fracton}
\end{figure}

We can also write down the defect that describes this motion. Let $i = i_0 \rightarrow i_1 \rightarrow \cdots \rightarrow i_{K-1} \rightarrow i_K = i'$ be the unique path from $i$ to $i'$.\footnote{Note that the arrows do not imply that the edges are directed; they only indicate that the path is from $i$ to $i'$.} The following defect describes the motion of a particle from $i$ to $i'$ at time $\hat \tau=\hat \tau_0$:
\ie\label{lapA-mobiledefectonpath}
\exp\left[ i \sum_{\hat \tau<\hat \tau_0} \mathcal A_\tau(\hat \tau,i) \right]\exp\left[ i \sum_{k=1}^{K} \sum_{j\in \Gamma^{(k)}}\mathcal A(\hat \tau_0,j) \right]\exp\left[ i \sum_{\hat \tau\ge\hat \tau_0} \mathcal A_\tau(\hat \tau,i') \right]~,
\fe
where $\Gamma^{(k)}$ is the \emph{connected component} of $i_k$ in $\Gamma \setminus \langle i_{k-1}, i_k\rangle$ for $1\le k\le K$. 

\item
On the other hand, if there are at least two paths from $i$ to $i'$, then $S_{i_0}(i) \ne S_{i_0}(i')$. Moreover, there is a group homomorphism $\psi:\Jac(\Gamma)\rightarrow U(1)$ such that $\psi(S_{i_0}(i)) \ne \psi(S_{i_0}(i'))$.\footnote{Let $g,h$ be two distinct elements in $\Jac(\Gamma) = \prod_{a<\mathsf N} \mathbb Z_{r_a}$. Any element of $\Jac(\Gamma)$ can be represented as an $(\mathsf N-1)$-tuple: $(s_1,\ldots,s_{\mathsf N-1})$ where $0\le s_a<r_a$. Say $a<\mathsf N$ is the first component at which $g$ and $h$ differ. Consider the group homomorphism $\psi:\Jac(\Gamma)\rightarrow U(1)$ given by $\psi:(s_1,\ldots,s_{\mathsf N-1}) \mapsto e^{2\pi i s_a/r_a}$. It is clear that $\psi(g) \ne \psi(h)$.} So, $\psi \circ S_{i_0}$ is a $U(1)$-valued discrete harmonic function that takes different values at $i$ and $i'$. It follows that a particle cannot move from $i$ to $i'$.
\end{itemize}

\begin{figure}[t]
\begin{center}
\includegraphics[scale=0.2]{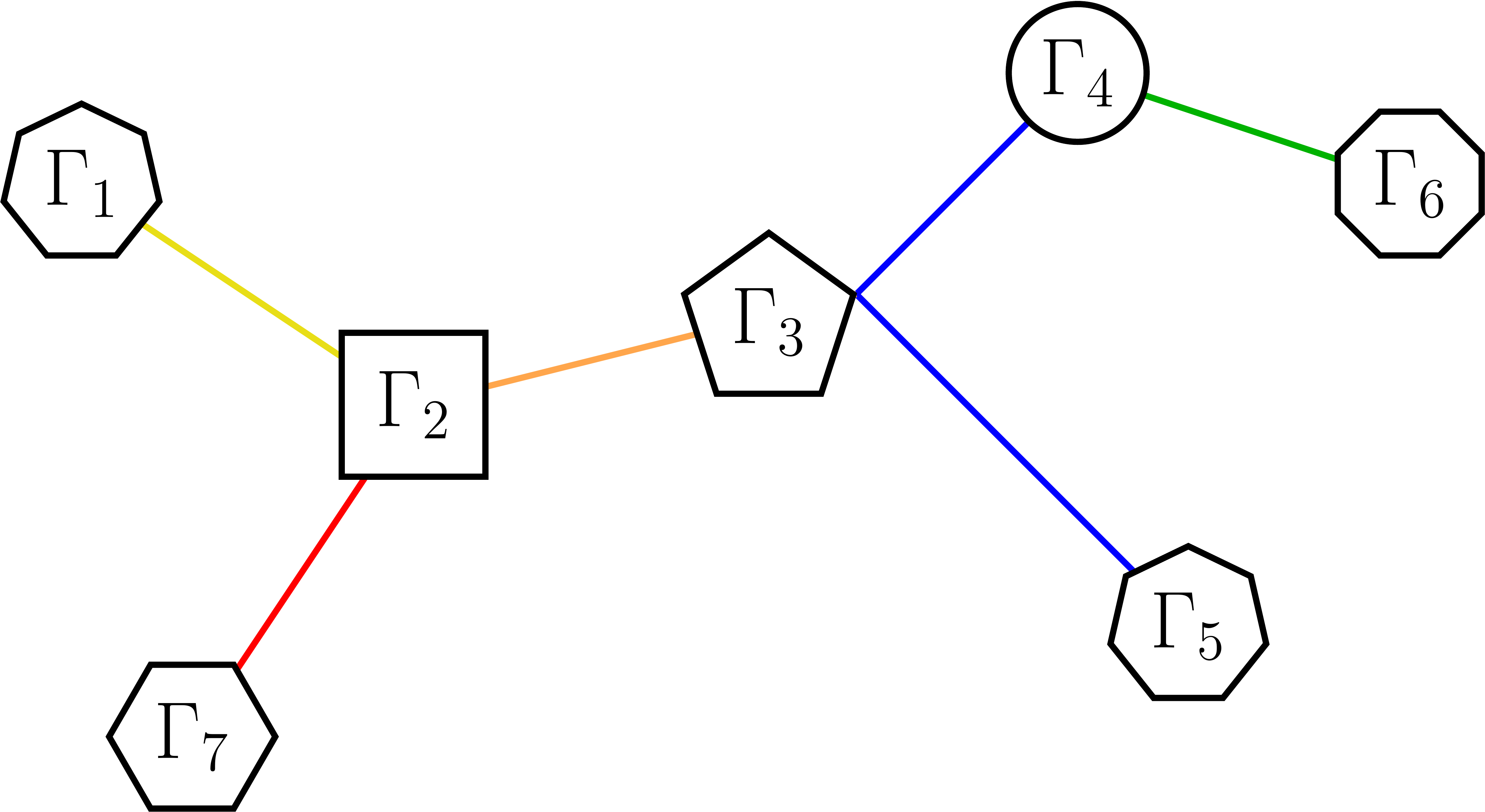}
\end{center}
\caption{The structure of any graph $\Gamma$, where each $\Gamma_k$ is a $2$-edge connected graph, and each line represents a path/bridge between two vertices from adjacent $\Gamma_k$'s. As explained in the main text, a particle on a bridge can move along that bridge or any other bridge that shares a vertex with that bridge but nowhere else, whereas a particle ``inside'' a $\Gamma_k$ cannot move anywhere. In the figure, this means, particles on colored parts of $\Gamma$ can move within the parts of same color.}\label{fig:graphstructure}
\end{figure}

In particular, if $\Gamma$ is $2$-edge connected (such as a torus graph), any particle on $\Gamma$ is a fracton. More generally, any $\Gamma$ has the structure shown in Figure \ref{fig:graphstructure}, where each $\Gamma_k$ is a $2$-edge connected graph, and the ``lines'' represent paths/bridges connecting two vertices from adjacent $\Gamma_k$'s---for example, when each $\Gamma_k$ is a single vertex, then $\Gamma$ is a tree. By the last two paragraphs, a particle on a bridge can move along that bridge or any other bridge that shares a vertex with that bridge but nowhere else, whereas a particle ``inside'' a $\Gamma_k$ cannot move anywhere.

\subsection{Correspondence between divisors and fractons}
In this subsection we unveil an  intriguing correspondence between the theory of divisors in graph theory and the $U(1)$ Laplacian gauge theory of fractons. A divisor $q\in\mathcal F(\Gamma,\mathbb Z)$ can be interpreted as a configuration of particles with the $U(1)$ time-like charge of the particle at a site $i$ equal to $q(i)\in \mathbb{Z}$. 
More specifically, a static configuration of particles associated with a divisor $q$ is represented by the  defect \eqref{divisorcorrespondence}. 
A configuration can ``move'' to another configuration if and only if   their corresponding divisors differ by a principal divisor. 
In particular, a configuration associated with a principal divisor can ``trivially move'' by first being annihilated and then being created elsewhere.

These statements can be understood by the selection rules imposed by the time-like symmetry. A time-like superselection sector is labelled by the time-like charges of a configuration. Since the time-like symmetry group is $U(1)\times \Jac(\Gamma)$, the time-like charges are valued in the Pontryagin dual, i.e., $\mathbb Z\times \Jac(\Gamma)$.\footnote{Here, the $\mathbb Z$ factor is the $U(1)$ time-like charge, which is associated with the degree of the divisor.} The latter is precisely the Picard group $\Pic(\Gamma)$. In other words, the superselection sector of a configuration is associated with the equivalence class of a divisor in $\Pic(\Gamma)$. In particular, the ``trivial'' superselection sector (with trivial time-like charges) corresponds to the equivalence class of $\Pic(\Gamma)$ that consists of all the principal divisors. Therefore, a configuration can ``move'' to another configuration if and only if they are in the same superselection sector, i.e., if and only if their divisors differ by a principal divisor   (see Figure \ref{fig:fracton} for an illustration).  This correspondence is summarized in Table \ref{tbl:div_fracton}.

\subsection{Examples}
In this section, we discuss the $U(1)$ Laplacian gauge theory on 1d and 2d spatial tori.

\subsubsection{1+1d rank-2 $U(1)$ tensor gauge theory}\label{sec:1ddipA}
Let $\Gamma$ be a \emph{cycle graph}, i.e., $\Gamma = C_{L_x} = \mathbb Z_{L_x}$, where $L_x$ is the number of sites in the cycle. The operator $\Delta_L$ associated with the Laplacian matrix of $\Gamma$ is the same as the standard Laplacian operator $\Delta_x^2$ in the $x$-direction. The action \eqref{lapA-modVill-action} becomes
\ie\label{1ddipA-modVill-action}
S = \frac{\gamma}{2} \sum_{\tau\text{-link}} \mathcal E_{xx}^2 + \frac{i\theta}{2\pi} \sum_{\tau\text{-link}} \mathcal E_{xx}~,
\fe
after replacing $\mathcal A$ with $\mathcal A_{xx}$. This is the modified Villain action of the 1+1d $U(1)$ dipole gauge theory or the rank-2 $U(1)$ tensor gauge theory \cite{Gorantla:2022eem}.

The Smith decomposition of $L$ is discussed in Section \ref{sec:1ddipphi}. It follows that the discrete electric time-like symmetry is $\mathbb Z_{L_x}$. Since $\Gamma$ is $2$-edge connected, a single particle cannot move, i.e., it is a fracton. On the other hand, dipoles can move (this is similar to the dipole motion shown in Figure \ref{fig:fracton}). This is in agreement with the analysis of \cite{Gorantla:2022eem}.

\subsubsection{2+1d $U(1)$ Laplacian gauge theory}
Let $\Gamma$ be a \emph{torus graph}, i.e., $\Gamma = C_{L_x} \times C_{L_y} = \mathbb Z_{L_x}\times \mathbb Z_{L_y}$, where $L_i$ is the number of sites in the $i$ direction. The operator $\Delta_L$ associated with the Laplacian matrix of $\Gamma$ is the same as the standard Laplacian operator $\Delta_x^2 + \Delta_y^2$ in the $xy$-plane. The action \eqref{lapA-modVill-action} becomes
\ie
S = \frac{\gamma}{2} \sum_{\tau\text{-link}} \mathcal E^2 + \frac{i\theta}{2\pi} \sum_{\tau\text{-link}} \mathcal E~,
\fe
where $\mathcal E = \Delta_\tau \mathcal A - (\Delta_x^2 + \Delta_y^2) \mathcal A_\tau - 2\pi m_\tau$. We refer to this as the 2+1d $U(1)$ Laplacian gauge theory. It is one of the natural higher dimensional versions of the 1+1d rank-2 $U(1)$ tensor gauge theory of Section \ref{sec:1ddipA}.

While the Jacobian group of a 2d torus graph is not generally known, since $\Gamma$ is $2$-edge connected, a single particle is a fracton. 
In fact, in an upcoming paper \cite{Gorantla:2022ssr}, we show that any finite set of particles with arbitrary charges is completely immobile (modulo sets of particles that can trivially move) on the infinite square lattice.

A closely related gauge theory is the 2+1d rank-2 $U(1)$ tensor gauge theory. While the 1+1d rank-2 $U(1)$ tensor gauge theory is the same as the 1+1d $U(1)$ Laplacian gauge theory, their 2+1d versions are very different. In an upcoming paper \cite{Gorantla:2022ssr}, we analyze both of them in detail, and compare their global properties.

\section*{Acknowledgements}

We would like to thank N. Seiberg for initial collaborations in this project and for insightful comments on a draft. 
We also thank  Y. Oz, T. Senthil, and A. Vishwanath for interesting discussions. PG would like to especially thank S. Velusamy for helpful discussions. PG was supported by Physics Department of Princeton University. HTL was supported in part by a Croucher fellowship from the Croucher Foundation, the Packard Foundation and the Center for Theoretical Physics at MIT. PG thanks the Department of Theoretical Physics at Tata Institute of Fundamental Research for its hospitality while this work was being completed. The authors of this paper were ordered alphabetically. Opinions and conclusions expressed here are those of the authors and do not necessarily reflect the views of funding agencies.

\appendix

\section{Discrete Poisson equation}\label{app:lap-poisson}
In this appendix, we solve the discrete Poisson equation \eqref{graph-discpoiss} using the Smith decomposition \eqref{graph-snf} of the Laplacian matrix.

Consider a function on the vertices of the graph, $f:\Gamma\rightarrow X$, where $X$ is an abelian group. We are interested in the case where $X$ is $\mathbb R$, $U(1)$, $\mathbb Z$, or $\mathbb Z_N$. Recall that the Laplacian operator $\Delta_L$ is defined as
\ie
\Delta_L f(i) = \sum_j L_{ij} f(j) = d_i f(i) - \sum_{j:\langle i,j\rangle\in \Gamma} f(j) = \sum_{j:\langle i,j\rangle\in \Gamma} [f(i) - f(j)]~.
\fe
One is usually interested in the following questions: given a function $g$, is there a function $f$ that satisfies the \emph{discrete Poisson equation} on the graph,
\ie\label{discrete-Poisson}
\Delta_L f(i) = g(i)~?
\fe
If yes, how many solutions are there, and what are they?

One way to answer these questions is to use the \emph{Smith normal form} of $L$ \cite{Lorenzini:08}. The Smith normal form of $L$ is given by $R = P L Q$, where $P,Q\in GL_{\mathsf N}(\mathbb Z)$, and $R = \diag(r_1,\ldots,r_{\mathsf N})$ is an $\mathsf N\times \mathsf N$ diagonal integer matrix with nonnegative diagonal entries such that $r_a | r_{a+1}$ (i.e., $r_a$ divides $r_{a+1}$) for $a=1,\ldots, \mathsf N-1$. While $R$ is uniquely determined by $L$, the matrices $P$ and $Q$ are not.

Using the index notation, we can write the Smith normal form as $R_{ab} = \sum_{i,j} P_{ai}L_{ij}Q_{jb}$, where $R_{ab} = r_a \delta_{ab}$. While all the indices run from $1$ to $\mathsf N$, the indices $i,j$ have natural interpretation as vertices of the graph $\Gamma$, whereas the indices $a,b$ do not have an immediately obvious interpretation.

For a connected graph $\Gamma$, we have $r_a>0$ for $a=1,\ldots, \mathsf N-1$, and $r_{\mathsf N}=0$. This implies
\ie\label{Pinv-Qinv-sum}
\sum_i (P^{-1})_{ia} = 0~,\qquad \sum_i (Q^{-1})_{ai} = 0~,
\fe
for all $a<\mathsf N$. In fact, a convenient choice of $P$ and $Q$ is given by the block matrices:
\ie
P = \begin{pmatrix}
	\tilde P & \b 0\\
	\b 1^T & 1
\end{pmatrix}~,\qquad Q = \begin{pmatrix}
	\tilde Q & \b 1\\
	\b 0^T & 1
\end{pmatrix}~,
\fe
where $\tilde P,\tilde Q\in GL_{\mathsf N-1}(\mathbb Z)$. It follows that
\ie
P^{-1} = \begin{pmatrix}
	\tilde P^{-1} & \b 0\\
	-\b 1^T\tilde P^{-1} & 1
\end{pmatrix}~,\qquad Q^{-1} = \begin{pmatrix}
	\tilde Q^{-1} & -\tilde Q^{-1}\b 1\\
	\b 0^T & 1
\end{pmatrix}~,
\fe
where \eqref{Pinv-Qinv-sum} is manifest.

The discrete Poisson equation \eqref{discrete-Poisson} simplifies in a new basis. Defining
\ie
&f'_a = \sum_i (Q^{-1})_{ai} f(i)\,,\\
&g'_a = \sum_i P_{ai} g(i),
\fe
 we can write \eqref{discrete-Poisson} as
\ie\label{disc-Poisson-ind}
r_a f'_a = g'_a~,\qquad a = 1,\ldots,\mathsf N~.
\fe
Each $a$ gives an independent equation, so we can solve for each $f'_a$ independently. If a solution exists for all $a$, then a solution exists for \eqref{discrete-Poisson}, and vice versa.

Let us first focus on $a=\mathsf N$. Since $r_{\mathsf N} = 0$, a solution exists if and only if $g'_{\mathsf N} = \sum_i P_{\mathsf N i} g(i) = \sum_i g(i) = 0$. 
This condition is the same for any choice of  $X$.
 If it is satisfied, then $f'_{\mathsf N}$ can take any value in $X$.\footnote{Note that since $Q\in GL_{\mathsf N}(\mathbb Z)$, if $f$ takes values in $X$, then $f'$ also takes values in $X$.} This corresponds to the zero mode (or constant mode because $Q_{i \mathsf N} = 1$ for all $i$) of the Laplacian.

Now, consider $a<\mathsf N$ so that $r_a > 0$. Let us analyze each $X$ separately.
\begin{itemize}

\item When $X = \mathbb R$, there is always a unique solution
\ie
f'_a = \frac{1}{r_a} g'_a~,
\fe
Assuming $g'_{\mathsf N} = 0$, the solution in the original basis is
\ie\label{Poisson-sol-R}
f(i) = \sum_{a< \mathsf N}\sum_j \frac{Q_{ia} P_{aj}}{r_a} g(j) + c~,
\fe
where $c$ is a real constant (it is the zero mode mentioned above).

\item When $X = U(1)$, the equation \eqref{disc-Poisson-ind} can be written as
\ie
r_a f'_a = g'_a \mod 2\pi~.
\fe
A general solution takes the form
\ie
f'_a = \frac{1}{r_a} g'_a + \frac{2\pi p_a}{r_a}~,
\fe
where $p_a$ is an integer. Since $X=U(1)$, the integers $p_a$ and $p_a + r_a$ correspond to the same solution, so there are $r_a$ inequivalent solutions associated with $p_a = 0,\ldots, r_a-1$.
Hence, assuming $g'_{\mathsf N} = 0 \mod 2\pi$, the general solution in the original basis is
\ie\label{Poisson-sol-U1}
f(i) = \sum_{a< \mathsf N}\sum_j \frac{Q_{ia} P_{aj}}{r_a} g(j) + 2\pi \sum_{a< \mathsf N} \frac{Q_{ia} p_a}{r_a} + c~,
\fe
where $c\sim c+2\pi$ is a circle-valued constant. If we lift $f$ and $g$ from $U(1)$ to $\mathbb R$, we have
\ie\label{Poisson-sol-U1-int}
\Delta_L f(i) = g(i) + 2\pi \sum_{a<\mathsf N} (P^{-1})_{ia} p_a~.
\fe

\item When $X = \mathbb Z$, a unique solution
\ie
f'_a = \frac{1}{r_a} g'_a~,
\fe
exists if and only if $r_a$ divides $g'_a$. 
So, assuming $g'_{\mathsf N} = 0$, and $g'_a = 0 \mod r_a$ for $a<\mathsf N$, the solution is
\ie
f(i) = \sum_{a< \mathsf N}\sum_j \frac{Q_{ia} P_{aj}}{r_a} g(j) + p~,
\fe
where $p$ is an integer.

\item When $X = \mathbb Z_N$, the equation \eqref{disc-Poisson-ind} can be written as
\ie
r_a f'_a = g'_a \mod N~.
\fe
A solution of the form
\ie
f'_a = \frac{\tilde r_a}{\gcd(N,r_a)} g'_a + \frac{N p_a}{\gcd(N,r_a)}~,
\fe
exists if and only if $\gcd(N,r_a)$ divides $g'_a$ due to B\'ezout's identity. Here, $\tilde r_a$ is a fixed integer given by $\tilde r_a r_a = \gcd(N,r_a) \mod N$ (which exists by B\'ezout's identity), and $p_a$ is an integer.\footnote{There are infinitely many choices of $\tilde r_a$ but they can be absorbed into $p_a$.} Note that $p_a \sim p_a + \gcd(N,r_a)$ because $X=\mathbb Z_N$, so there are $\gcd(N,r_a)$ inequivalent solutions associated with $p_a = 0,\ldots, \gcd(N,r_a)-1$. So, assuming $g'_{\mathsf N} = 0 \mod N$, and $g'_a = 0 \mod \gcd(N,r_a)$ for $a<\mathsf N$, the general solution is
\ie
f(i) = \sum_{a< \mathsf N}\sum_j \frac{Q_{ia} \tilde r_a P_{aj}}{\gcd(N,r_a)} g(j) + \sum_{a< \mathsf N} \frac{NQ_{ia} p_a}{\gcd(N,r_a)} + p~,
\fe
where $p$ is an integer modulo $N$. Since $\gcd(N,r_{\mathsf N}) = \gcd(N,0) = N$, and $Q_{i\mathsf N}=1$, defining $p_{\mathsf N} = p \mod N$, we can write the above solution as
\ie\label{Poisson-sol-ZN}
f(i) = \sum_{a< \mathsf N}\sum_j \frac{Q_{ia} \tilde r_a P_{aj}}{\gcd(N,r_a)} g(j) + \sum_a \frac{NQ_{ia} p_a}{\gcd(N,r_a)}~,
\fe
which exists if and only if $g'_a=0\mod \gcd(N,r_a)$ for all $a$.

\end{itemize}

There is another useful perspective to the above analysis. Say $g$ is a gauge field on $\Gamma$, and $f$ is its gauge parameter with gauge symmetry
\ie
g(i) \sim g(i) - \Delta_L f(i)~.
\fe
In other words, $g$ represents an equivalence class in $\coker_X \Delta_L$. We are interested in the gauge invariant information in $g$, i.e., the \emph{holonomies} of $g$. Alternatively, we can ask what part of $g$ can be gauge-fixed to zero, i.e., if there is an $f$ satisfying \eqref{discrete-Poisson}. From the above analysis, the answer is the following:
\begin{itemize}

\item When $X=\mathbb R$, the only holonomy of $g$ is $g'_{\mathsf N} = \sum_i g(i)$. If $g'_{\mathsf N} = c$, then we can use the gauge freedom to gauge-fix $g$ to be of the form
\ie
g(i) = (P^{-1})_{i\mathsf N} g'_{\mathsf N} = \delta_{i\mathsf N} c~,\qquad \text{or} \qquad g(i) = \frac{c}{\mathsf N}~.
\fe
Here, $c$ is a real constant.

\item When $X=U(1)$, the only holonomy of $g$ is $g'_{\mathsf N} = \sum_i g(i) \mod 2\pi$. If $g'_{\mathsf N} = c$, then we can use the gauge freedom to gauge-fix $g$ to be of the form
\ie\label{gaugefix-U1}
g(i) = (P^{-1})_{i\mathsf N} g'_{\mathsf N} = \delta_{i\mathsf N} c~,\qquad \text{or} \qquad g(i) = \frac{c}{\mathsf N}~.
\fe
Here, $c\sim c+2\pi$ is a circle-valued constant.

\item When $X=\mathbb Z$, the holonomies of $g$ are $g'_{\mathsf N} = \sum_i g(i)$, and $g'_a = \sum_i P_{ai} g(i) \mod r_a$ for $a<\mathsf N$. If $g'_{\mathsf N} = p_{\mathsf N}$ and $g'_a = p_a \mod r_a$, then we can use the gauge freedom to gauge-fix $g$ to be of the form
\ie\label{gaugefix-Z}
g(i) = \sum_a (P^{-1})_{ia} p_a~.
\fe
Here, $p_{\mathsf N}$ is an integer, and $p_a = 0,\ldots,r_a-1$ for $a<\mathsf N$. When $p_{\mathsf N} = 0$, each function in \eqref{gaugefix-Z} represents a unique equivalence class of $\Jac(\Gamma)$. Therefore, 
\ie
\Jac(\Gamma) \cong \prod_{a<\mathsf N} \mathbb Z_{r_a}~.
\fe

\item When $X=\mathbb Z_N$, the holonomies of $g$ are $g'_{\mathsf N} = \sum_i g(i) \mod N$, and $g'_a = \sum_i P_{ai} g(i) \mod \gcd(N,r_a)$ for $a<\mathsf N$. Since $r_{\mathsf N} = 0$, we can combine them into $g'_a = \sum_i P_{ai} g(i) \mod \gcd(N,r_a)$ for all $a$. If $g'_a = p_a \mod \gcd(N,r_a)$, then we can use the gauge freedom to gauge-fix $g$ to be of the form
\ie\label{gaugefix-ZN}
g(i) = \sum_a (P^{-1})_{ia} p_a~.
\fe
Here, $p_a = 0,\ldots,\gcd(N,r_a)-1$ for all $a$.

\end{itemize}

\bibliographystyle{JHEP}
\bibliography{laplacian,fracton}

\end{document}